\documentclass[preprint,12pt]{elsarticle}

\usepackage{amssymb}
\usepackage{amsmath}  %% For extended equation environments
\usepackage{graphicx} %% For including figures
\usepackage{hyperref} %% For hyperlinks in references (optional)
\usepackage{verbatim}
\usepackage{algorithm}
\usepackage{url} % For the \url command
\PassOptionsToPackage{hyphens}{url}
\usepackage{hyperref}
\usepackage{colortbl} % For coloring rows
\usepackage{booktabs}
\usepackage{xcolor}   % For color definitions
\usepackage[noend]{algpseudocode}

\journal{Applied Soft Computing}

\begin{document}
\begin{frontmatter}

\title{LENS-XAI: Redefining Lightweight and Explainable Network Security through Knowledge Distillation and Variational Autoencoders for Scalable Intrusion Detection in Cybersecurity}

\author[label1]{Muhammet Anil Yagiz}
\affiliation[label1]{organization={Department of Computer Engineering, Kırıkkale University}, 
            city={Kırıkkale}, 
            postcode={71450}, 
            country={Turkey}}

\author[label2]{Polat Goktas\corref{cor1}}
\affiliation[label2]{organization={School of Computer Science \& CeADAR, University College Dublin}, 
            city={Dublin}, 
            postcode={D04 V2N9}, , 
            country={Ireland}}

%% Corresponding author information
\cortext[cor1]{Corresponding author. Email: polat.goktas@ucd.ie}

%% Abstract
\begin{abstract}
The rapid proliferation of Industrial Internet of Things (IIoT) systems necessitates advanced, interpretable, and scalable intrusion detection systems (IDS) to combat emerging cyber threats. Traditional IDS face challenges such as high computational demands, limited explainability, and inflexibility against evolving attack patterns. To address these limitations, this study introduces the Lightweight Explainable Network Security framework (LENS-XAI), which combines robust intrusion detection with enhanced interpretability and scalability. LENS-XAI integrates knowledge distillation, variational autoencoder models, and attribution-based explainability techniques to achieve high detection accuracy and transparency in decision-making. By leveraging a training set comprising  $10\%$ of the available data, the framework optimizes computational efficiency without sacrificing performance. 
Experimental evaluation on four benchmark datasets—Edge-IIoTset, UKM-IDS20, CTU-13, and NSL-KDD—demonstrates the framework’s superior performance, achieving detection accuracies of 95.34\%, 99.92\%, 98.42\%, and 99.34\%, respectively. Additionally, the framework excels in reducing false positives and adapting to complex attack scenarios, outperforming existing state-of-the-art methods. Key strengths of LENS-XAI include its lightweight design, suitable for resource-constrained environments, and its scalability across diverse IIoT and cybersecurity contexts. Moreover, the explainability module enhances trust and transparency, critical for practical deployment in dynamic and sensitive applications. This research contributes significantly to advancing IDS by addressing computational efficiency, feature interpretability, and real-world applicability. Future work could focus on extending the framework to ensemble AI systems for distributed environments, further enhancing its robustness and adaptability.
\end{abstract}

%% Keywords

\end{frontmatter}

\section{Introduction}
\label{sec:1-introduction}
In the ever-expanding digital landscape, the rapid proliferation of interconnected devices has led to a corresponding rise in sophisticated cyber threats, challenging the efficacy of traditional network security measures \cite{alrashdi2024fidwatch}. Intrusion Detection Systems (IDS) have emerged as critical components in safeguarding networks by identifying and mitigating anomalous activities \cite{ullah2024idsint}. However, the increasing complexity of cyberattacks necessitates the adoption of advanced methodologies that can keep pace with evolving threats while maintaining interpretability and scalability \cite{nkoro2024zerotrust}. Recent studies highlights the adoption of advanced methodologies, including the integration of deep learning (DL) techniques with artifical intelligence (AI) for enhanced intrusion detection \cite{fatema2025securing, kumar2024}, federated incremental learning for Internet of Things (IoT) security monitoring \cite{alrashdi2024fidwatch}, and blockchain-enhanced decision-making frameworks \cite{kumar2024}. Furthermore, intelligent systems tailored for specialized environments like unmanned aerial vehicles highlight the potential of interpretable IDS in addressing domain-specific challenges \cite{javeed2024intelligent}.

Despite this progress, a significant challenge persists: the \textit{``black-box"} nature of these models undermines their adoption in critical applications where explainability and transparency are paramount. Addressing this gap, Explainable AI (XAI) has gained prominence, enabling researchers and practitioners to understand and trust IDS based on machine learning (ML) / DL \cite{xai_intro, xai_trust}. Incorporating XAI frameworks into IDS improves both operational transparency and decision-making reliability. For example, models such as DeepRoughNetID \cite{nalini2024deeproughnetid} have demonstrated robust anomaly detection capabilities by combining feature engineering with interpretable algorithms. Similarly, the integration of XAI into IoT-based intrusion detection systems, as illustrated by TwinSec-IDS \cite{krishnaveni2024twinsec}, highlights the potential of explainable frameworks in addressing security challenges in complex environments. These advancements underline the necessity of pairing state-of-the-art ML/DL techniques with XAI to foster adoption and trust in IDS solutions.

The contemporary cybersecurity ecosystem also demands lightweight models that can operate efficiently in resource-constrained settings such as IoT and edge computing environments. For example, the Zero-Trust Marine Cyberdefense framework \cite{nkoro2024zerotrust} exemplifies the integration of lightweight architectures with explainable mechanisms, showcasing their effectiveness in specialized domains. This approach aligns with industry trends that prioritize the dual objectives of computational efficiency and interpretability. Among the cutting-edge methodologies in IDS development, knowledge distillation and variational autoencoders (VAEs) have emerged as transformative techniques. Knowledge distillation facilitates the transfer of insights from complex, high-capacity models to compact, efficient ones, ensuring that resource-efficient systems do not compromise on performance \cite{kd_intro}. Meanwhile, VAEs have proven instrumental in learning latent representations of data, allowing anomaly detection by analyzing deviations from normal patterns \cite{vae_intro}. Leveraging these methodologies within an XAI framework can redefine the paradigms of network security.

Despite these advances, a critical gap persists in the current literature. Although existing approaches address aspects of either scalability, interpretability, or computational efficiency, comprehensive solutions that holistically integrate these factors remain scarce. In particular, studies such as PANACEA \cite{panacea} and XAIEnsembleTL-IoV \cite{xai_ensemble_iov} emphasize specific facets of IDS but fail to provide an integrated framework applicable in diverse real-world scenarios, particularly in resource-constrained environments. Furthermore, the absence of robust explanations for model predictions often hinders the deployment of these systems in critical applications where trust and accountability are paramount. In the evolving landscape of autonomous vehicles, ensuring robust in-vehicle network security is paramount. 

In this study, we introduce \textbf{LENS-XAI}, an advanced framework that redefines lightweight and explainable network security by combining knowledge distillation with VAEs. The LENS-XAI approach is tailored for resource-constrained environments, such as IoT and edge computing, emphasizing scalability and interpretability. Its well-designed architecture enables high detection accuracy, robust anomaly detection, and actionable insights into decision-making processes. This framework seeks to balance performance and transparency, offering an effective solution for intrusion detection in modern network infrastructures. The primary contributions of this study include:

\begin{itemize}
    \item \textbf{Innovative Integration of Knowledge Distillation and VAEs:} The framework employs knowledge distillation for enhanced computational efficiency and VAEs to model complex data distributions, improving detection capabilities.
    \item \textbf{Enhanced Explainability:} LENS-XAI incorporates variable attribu-tion-based XAI mechanisms mechanisms to provide detailed insights, fostering greater trust and clarity in its operations.
    \item \textbf{Optimized for Resource-Constrained Environments:} Its scalable and lightweight architecture is designed to adapt seamlessly to IoT and edge computing scenarios without sacrificing detection performance.
    \item \textbf{Robust Validation:} Extensive experiments on benchmark datasets, including NSL-KDD, CTU-13, UKM20 and EDGEIIoTset, demonstrate the framework's efficacy and reliability in diverse real-world applications.
\end{itemize}
The remainder of this paper is organized as follows. Section II reviews related works in ML/DL-based IDS and XAI integration. Section III outlines the methodology, detailing the proposed model architecture and its components. Section IV presents experimental results, highlighting performance metrics and comparative analyses. Finally, Section V \& VI conclude with insights and future directions for advancing IDS research.

\section{Literature survey}
\label{sec:2-literature_survey}
\label{sec:2-literature_survey}
IDSs have been widely researched to address the increasing sophistication of cyberattacks. This section reviews recent advances in IDS methodologies with an emphasis on using XAI and lightweight architectures to enhance interpretability and efficiency.

\subsection{Explainable Deep Learning Approaches for IDS}
DL models have dramatically transformed IDS by providing enhanced accuracy and adaptability in identifying malicious activities. However, their inherent \textit{``black-box"} nature raises significant concerns about trust and interpretability. To address these issues, Sindiramutty et al. \cite{sindiramutty2024} proposed a Bi-LSTM-based framework integrated with explainable mechanisms, including enhanced krill herd optimization, which provides valuable insights into its decision-making process. This framework has proven effective in industrial cyber-physical systems, emphasizing the need for models that balance interpretability with high performance. Similarly, Nalini et al. \cite{nalini2024deeproughnetid} introduced DeepRoughNetID, a model designed to enhance anomaly detection capabilities by combining robust feature engineering with interpretable algorithms. While achieving high detection rates, DeepRoughNetID highlights the importance of evaluating its effectiveness in resource-constrained environments, a domain where its applicability remains underexplored. Al-Essa et al. \cite{panacea} extended this line of research by developing PANACEA, a neural model ensemble utilizing knowledge distillation to enhance both efficiency and explainability. While PANACEA demonstrates its potential through accurate detection rates, its reliance on specific datasets limits its generalizability across diverse network environments. Alotaibi et al. \cite{alotaibi2025} applied XAI to web phishing classification in IoT and cyber-physical systems, achieving strong results but requiring further validation across diverse IoT architectures.  Zhao et al. \cite{zhao2021} also addressed these challenges by proposing a lightweight IDS framework based on knowledge distillation and deep metric learning. However, their approach fell short in addressing the few-shot learning problem, leaving room for further optimization.

\subsection{Lightweight and Scalable Models}
As IoT devices and edge computing architectures continue to proliferate, IDS models must adapt to operate effectively in resource-constrained settings. Lightweight and scalable frameworks have become essential in addressing these demands. Nkoro et al. \cite{nkoro2024zerotrust} proposed the Zero-Trust Marine Cyberdefense framework, which integrates lightweight architectures with explainable mechanisms to ensure robust cybersecurity in IoT-based maritime networks. This approach exemplifies the effectiveness of models designed for specialized environments, emphasizing computational efficiency without compromising interpretability. Ullah et al. \cite{ullah2024idsint} presented IDS-INT, a transformer-based intrusion detection model that employs transfer learning to handle unbalanced network traffic. While its advanced architecture achieves high detection accuracy, the model's scalability across different network configurations remains a critical area for further validation. 

Chen et al. \cite{chen2023} explored federated learning for IDS, incorporating differentially private knowledge distillation to preserve data privacy while enhancing classification accuracy. Their framework balanced privacy and utility, but required optimization for heterogeneous IoT architectures. Similarly, Kumar et al. \cite{kumar2024} developed a blockchain-enabled IDS integrated with XAI for enhanced decision-making transparency in industrial systems. While promising, their approach demands further evaluation in dynamic industrial contexts. In parallel, Fatema et al. \cite{fatema2025securing} introduced a federated learning framework combined with XAI techniques to enhance the security and efficiency of IDS in IoT ecosystems. Their work highlights the growing need for collaborative and distributed approaches that ensure scalability and trustworthiness in real-world applications.

\subsection{Advanced Anomaly Detection Techniques}
Modern anomaly detection methods emphasize the need to capture complex data patterns and deviations effectively, particularly in dynamic network environments. Gaspar et al. \cite{xai_trust} explored the integration of SHAP (SHapley Additive exPlanations) and LIME (Local Interpretable Model-Agnostic Explanations) techniques in intrusion detection systems based on Multi-Layer Perceptron models. By combining local and global interpretability, their approach not only improves detection accuracy but also enhances the transparency of model predictions, fostering trust among users. Bacevicius et al. \cite{bacevicius2023mlrawdata} investigated hybrid approaches to handle unbalanced intrusion detection datasets, demonstrating improved classification accuracy in multi-class scenarios. Roy et al. \cite{roy2022trustworthy} proposed an explainable deep neural framework tailored for industrial settings, enabling trustworthy and transparent anomaly detection processes. In addition, Ahmed et al. \cite{ahmed2024} developed a hybrid ensemble IDS model incorporating bagging, boosting, and SHAP to achieve high accuracy. These methodologies underline the importance of combining interpretability with advanced learning mechanisms to address the complexities of modern cyber threats.

\subsection{Knowledge Distillation \& VAEs}
Knowledge distillation and VAEs have emerged as powerful tools for building efficient and interpretable IDS frameworks. By transferring knowledge from high-capacity models to lightweight counterparts, knowledge distillation ensures that computationally efficient systems maintain high detection accuracy. Sindiramutty et al. \cite{sindiramutty2024} effectively combined these techniques to design an anomaly detection model capable of balancing efficiency and explainability in industrial applications. Similarly, Moustafa et al. \cite{moustafa2023iotdefense} integrated VAEs with federated learning and XAI frameworks to develop a robust intrusion detection solution for IoT networks. Javeed et al. \cite{javeed2024intelligent} further demonstrated the utility of these approaches in specialized domains, designing an interpretable IDS tailored for unmanned aerial vehicles. 

Building on prior advancements, our earlier study by Yagiz et al. \cite{yagiz2024transforming} introduced the KD-XVAE system, integrating a Knowledge Distillation framework with VAEs. This model achieved exceptional performance, with perfect recall, precision, and F1 scores, while maintaining a lightweight computational design. By employing XAI techniques such as SHAP, the KD-XVAE system enabled interpretability by highlighting key latent features critical for decision-making. These studies highlight the transformative potential of integrating knowledge distillation and VAEs into XAI-based intrusion detection systems, paving the way for innovative solutions that meet the demands of modern network security.

\subsection{Research Gap}
The reviewed works highlight several research gaps in IDS development. Bacevicius et al. \cite{bacevicius2023mlrawdata} and Le et al. \cite{Le2023} emphasized the challenges associated with raw and unbalanced intrusion detection datasets, particularly in multi-class classification problems. Moustafa et al. \cite{moustafa2023iotdefense} and Kostopoulos et al. \cite{Kostopoulos2023} explored the use of XAI to address the interpretability issues of IDS models. Furthermore, Shtayat et al. \cite{shtayat2023} and Sivamohan et al. \cite{sivamohan2023} demonstrated the potential of integrating ensemble learning and DL techniques for robust intrusion detection but noted limitations in scalability across diverse IoT environments. The computational challenges and lack of interpretability in intrusion detection frameworks were also highlighted in works by Hattak et al. \cite{Hattak2023} and Arisdakessian et al. \cite{arisdakessian2023}. Additionally, the need for enhanced attack classification and explanations in dynamic IoT environments was underlined by Gaspar et al. \cite{xai_trust} and Kumar et al. \cite{kumar2024a}. .

In response to these gaps, our proposed \textbf{LENS-XAI} framework integrates knowledge distillation and VAEs within a lightweight architecture, tailored for resource-constrained IoT and edge computing environments. Emphasizing scalability, interpretability, and efficiency, it employs variable \linebreak attribution-based XAI mechanisms to provide transparent insights into \linebreak decision-making processes. By validating its performance across diverse datasets and scenarios, this framework advances the state-of-the-art in IDS.

\subsection{Problem Statement and Motivation}
Existing IDS frameworks face significant challenges in scalability, dataset-specific performance, and adaptability to dynamic environments. Many models rely on static feature selection techniques and heuristic-driven metrics, which limit their ability to identify critical relationships among features, particularly in the context of imbalanced datasets. Additionally, the computational overhead of existing solutions restricts their real-time applicability, especially in resource-constrained environments like IoT and edge computing. The limited scenario-based evaluations and reliance on specific datasets further constrain the generalizability of these models, leaving critical gaps in their ability to address diverse real-world applications. Moreover, the absence of dynamic threat adaptation mechanisms undermines the effectiveness of IDS frameworks in handling evolving cyber threats.
These challenges necessitate the development of robust, scalable, and generalizable IDS solutions capable of operating efficiently in resource-constrained environments. By addressing these limitations, it becomes possible to enhance the overall efficiency, adaptability, and reliability of IDS frameworks, ensuring improved security across diverse and dynamic network scenarios.

\section{Methodology}
\label{sec:3-methodology}
%The proposed methodology introduces a comprehensive IDS framework leveraging advanced data pre-processing techniques, VAEs, and Knowledge Distillation, enhanced by interpretability through a variable attribution-based explainability approach.  
%The overall process is outlined in Fig. 1.

\subsection{Framework Overview}
The proposed workflow begins with input datasets, which undergo pre-processing to remove inconsistencies, transform categorical attributes, and normalize numerical features. The processed data is fed into a VAE, capturing latent representations critical for anomaly detection. To enhance efficiency, Knowledge Distillation transfers the learned representations from a robust teacher model to a lightweight student model, optimizing computational performance for resource-constrained environments. Finally, a variable attribution-based explainability method provides transparent insights into the decision-making process, enhancing trust and usability in the detection framework.

\subsection{Data Preprocessing and Feature Engineering}

Real-world intrusion detection datasets (\textit{e.g.}, CTU-13, UKM20, NSL-KDD, and Edge-IIoTset) often contain missing values, categorical features, and different numerical scales. To ensure consistency and integrity, we perform three key preprocessing steps: \textit{handling missing values}, \textit{encoding categorical variables}, and \textit{normalizing continuous features}.

\subsubsection{Missing Value Handling}
We adopt a hybrid strategy to address missing entries:
\begin{enumerate}
    \item \textbf{Imputation:} For minor gaps, we replace missing values with the mean or median, preserving overall statistical distributions.
    \item \textbf{Exclusion:} For records with extensive missingness, we remove them to minimize noise and bias.
\end{enumerate}
This approach retains the majority of the data while limiting the distortion introduced by imputation.

\subsubsection{Feature Encoding and Transformation}
Categorical and non-numerical features must be translated into numerical forms for downstream ML models:
\begin{itemize}
    \item \textbf{One-hot encoding:} Converts nominal categories into binary indicator variables.
    \item \textbf{Ordinal encoding:} Maps ordered categories to integer values when a natural order is relevant.
\end{itemize}
We choose the encoding method on a feature-by-feature basis, prioritizing minimal information loss.

\subsubsection{Standardization}
To standardize the range of numerical attributes, we employ Z-score standardization. Each feature \( x \) is scaled to \( x' \) via
\begin{equation}
x' = \frac{x - \mu}{\sigma}
\end{equation}
where \( \mu \) and \( \sigma \) are the mean and standard deviation of the feature, respectively. This standardization ensures that the features have a mean of 0 and a standard deviation of 1, facilitating balanced gradient updates during model training.

\subsection{Representation Learning with Variational Autoencoders}

Following preprocessing, we employ VAEs to model the underlying data distribution and derive expressive latent representations that facilitate anomaly detection.

\subsubsection{VAE Architecture}
A standard VAE comprises two main components:
\begin{itemize}
    \item \textbf{Encoder:} A neural network $q_\phi(\mathbf{z} \mid \mathbf{x})$ that maps the input $\mathbf{x} \in \mathbb{R}^d$ to a latent distribution over $\mathbf{z} \in \mathbb{R}^k$.
    \item \textbf{Decoder:} A neural network $p_\theta(\mathbf{x} \mid \mathbf{z})$ that reconstructs $\mathbf{x}$ from latent variables $\mathbf{z}$.
\end{itemize}
Here, $\phi$ and $\theta$ denote the trainable parameters of the encoder and decoder, respectively, while $k$ is typically chosen to be much smaller than $d$.

\subsubsection{VAE Loss Formulation}
We optimize the VAE via the following objective:
\begin{equation}
\mathcal{L}_{\text{VAE}}(\theta, \phi; \mathbf{x}) 
= \underbrace{\mathbb{E}_{q_\phi(\mathbf{z} \mid \mathbf{x})} 
\big[ - \log p_\theta(\mathbf{x} \mid \mathbf{z}) \big]}_{\text{Reconstruction loss}} 
\;+\; \beta \underbrace{D_{\text{KL}}\big(q_\phi(\mathbf{z} \mid \mathbf{x}) \;\|\; p(\mathbf{z})\big)}_{\text{KL divergence}}
\end{equation}
where:
\begin{itemize}
    \item $\mathbb{E}_{q_\phi(\mathbf{z} \mid \mathbf{x})} \big[ - \log p_\theta(\mathbf{x} \mid \mathbf{z}) \big]$ measures the reconstruction error,
    \item $D_{\text{KL}}\big(q_\phi(\mathbf{z} \mid \mathbf{x}) \,\|\, p(\mathbf{z})\big)$ enforces a structured latent space by penalizing divergence from the prior $p(\mathbf{z})$, often chosen as $\mathcal{N}(\mathbf{0}, \mathbf{I})$,
    \item $\beta$ balances the trade-off between reconstruction fidelity and latent-space regularization~\cite{kingma2014auto}.
\end{itemize}

\subsubsection{Latent Representation for Anomaly Detection}
We use the encoder’s learned representations to identify anomalous samples in subsequent classification or threshold-based detection steps. Intuitively, examples that yield high reconstruction error or have low likelihood under the learned latent distribution are flagged as potential intrusions.

\subsection{Knowledge Distillation for Model Optimization}

VAEs combined with high-capacity classification models can be computationally demanding. To alleviate inference costs without sacrificing performance, we employ Knowledge Distillation~\cite{hinton2015distilling}, transferring the \textit{``knowledge''} from a powerful teacher model to a more compact student model.

\subsubsection{Distillation Setup}
\begin{enumerate}
    \item \textbf{Teacher Model:} Trained on the full dataset (or its latent representations) to achieve high accuracy.
    \item \textbf{Student Model:} A lighter network optimized to reproduce the output distribution of the teacher.
\end{enumerate}

\subsubsection{Distillation Loss}
To guide the student network, the teacher provides ``soft targets,'' which we combine with hard labels to form the distillation loss:
\begin{equation}
\mathcal{L}_{\text{distill}}(\theta_s) 
= (1-\alpha) \underbrace{\mathcal{L}_{\text{CE}}(y, \hat{y}_{s})}_{\text{hard label}} 
\;+\; \alpha T^2 \underbrace{D_{\text{KL}}\big(\sigma_T(\hat{y}_{t}), \,\sigma_T(\hat{y}_{s})\big)}_{\text{soft label}}
\end{equation}
where:
\begin{itemize}
    \item $\mathcal{L}_{\text{CE}}$ is cross-entropy between true labels $y$ and student predictions $\hat{y}_s$,
    \item $D_{\text{KL}}\big(\sigma_T(\hat{y}_{t}), \sigma_T(\hat{y}_{s})\big)$ is the KL divergence between teacher and student outputs under a ``softmax temperature'' $T$,
    \item $\alpha$ weights the relative contribution of hard and soft labels.
\end{itemize}
Here, $\sigma_T(\cdot)$ denotes the temperature-scaled softmax operator~\cite{hinton2015distilling}.

\subsection{Variable Attribution-Based Explainability}

Security analysts require interpretable results to validate and trust IDS outcomes. To meet this need, we incorporate a variable attribution-based explainability approach, as detailed in \textbf{Algorithm~\ref{alg:variable_attribution}}, which quantifies the contributions of individual features in anomaly predictions \cite{gosiewska2019ibreakdown, robniksikonja2008explaining}. This algorithm computes feature attributions by decomposing the model prediction for a given test instance into baseline and individual feature contributions, ensuring the property of \textit{local accuracy}. Specifically, the algorithm iteratively calculates the marginal contribution of each feature by comparing the expected model prediction conditioned on subsets of features. The process guarantees that the sum of all contributions aligns with the model's output, providing a transparent and interpretable breakdown of the prediction \cite{robniksikonja2018explainprediction, biecek2020explore}.

\subsubsection{Variable Attribution Computation}
Given an instance \(\mathbf{x}^* = \{x_1^*, x_2^*, \dots, x_p^*\}\), the model prediction \(f(\mathbf{x}^*)\) is expressed as \cite{robniksikonja2018explainprediction, biecek2020explore}:
\begin{equation}
f(\mathbf{x}^*) = v_0 + \sum_{j=1}^p v(j, \mathbf{x}^*),
\end{equation}
\vspace{-0.5em}
where:
\begin{itemize}
    \item \(v_0\) denotes the mean model prediction across all instances (baseline prediction),
    \item \(v(j, \mathbf{x}^*)\) represents the contribution of the \(j\)-th variable to the prediction \(f(\mathbf{x}^*)\).
\end{itemize}
This decomposition ensures the property of \textit{local accuracy}, such that:
\begin{equation}
\sum_{j=0}^p v(j, \mathbf{x}^*) = f(\mathbf{x}^*).
\end{equation}

To compute the contribution \(v(j, \mathbf{x}^*)\), we leverage conditional expectations:
\begin{equation}
v(j, \mathbf{x}^*) = \mathbb{E}\left[f(\mathbf{x}) \mid x_1 = x_1^*, \dots, x_j = x_j^*\right] - \mathbb{E}\left[f(\mathbf{x}) \mid x_1 = x_1^*, \dots, x_{j-1} = x_{j-1}^*\right],
\end{equation}
where:
\begin{itemize}
    \item \(\mathbb{E}[f(\mathbf{x}) \mid x_1, \dots, x_j]\) is the expected value of the model's prediction when the first \(j\) features are fixed to their observed values,
    \item The difference isolates the marginal contribution of the \(j\)-th variable.
\end{itemize}

\subsubsection{Generalization to Subsets of Features}

To handle more general cases, let \(J = \{j_1, j_2, \dots, j_K\}\) be a subset of \(K \leq p\) indices from \(\{1, 2, \dots, p\}\), and let \(L = \{l_1, l_2, \dots, l_M\}\) be another subset of \(M \leq p - K\) indices, such that \(J \cap L = \emptyset\). Define the conditional difference for \(L \mid J\) as \cite{biecek2020explore}:
\begin{equation}
\Delta_{L \mid J}(\mathbf{x}^*) = \mathbb{E}\left[f(\mathbf{x}) \mid \mathbf{x}_J = \mathbf{x}_J^*, \mathbf{x}_L = \mathbf{x}_L^*\right] - \mathbb{E}\left[f(\mathbf{x}) \mid \mathbf{x}_J = \mathbf{x}_J^*\right],
\end{equation}
where:
\begin{itemize}
    \item \(\mathbf{x}_J\) and \(\mathbf{x}_L\) denote the sets of variables indexed by \(J\) and \(L\), respectively,
    \item \(\mathbf{x}_J^*\) and \(\mathbf{x}_L^*\) represent their observed values.
\end{itemize}

For an individual variable \(l \in L\), the marginal contribution is:
\begin{equation}
\Delta_{l \mid J}(\mathbf{x}^*) = \mathbb{E}\left[f(\mathbf{x}) \mid \mathbf{x}_J = \mathbf{x}_J^*, x_l = x_l^*\right] - \mathbb{E}\left[f(\mathbf{x}) \mid \mathbf{x}_J = \mathbf{x}_J^*\right].
\end{equation}

\subsubsection{Order Sensitivity and Approximation}

The contributions \(v(j, \mathbf{x}^*)\) depend on the order of the variables \(j\), as interactions between variables may influence the result. To address this, we use a two-step heuristic \cite{biecek2020explore, breakdown2024}:
\begin{enumerate}
    \item Variables are ordered based on their marginal importance \(|\Delta_{j \mid \emptyset}(\mathbf{x}^*)|\), calculated without prior conditioning.
    \item Contributions \(v(j, \mathbf{x}^*)\) are then computed sequentially using the chosen order, capturing both individual and interaction effects.
\end{enumerate}

\begin{algorithm}[h]
\caption{Variable Attribution-Based Explainability Computation}
\label{alg:variable_attribution}
\begin{algorithmic}[1]
\Require $f$: Predictive model, $\mathbf{x}_{\text{test}}$: Test sample, $p$: Number of features
\Ensure Feature contributions $\{v(i, \mathbf{x}_{\text{test}})\}_{i=1}^p$
\State \textbf{Initialize:}
\Statex \hspace{\algorithmicindent} Baseline prediction $v_0 \gets \mathbb{E}[f(\mathbf{x})]$

\For{$i = 1$ to $p$}
    \State Compute conditional expectation for fixed features:
    \[
    \mathbb{E}[f(\mathbf{x}) \mid x_1, \dots, x_i]
    \]
    \State Compute conditional expectation for the previous subset:
    \[
    \mathbb{E}[f(\mathbf{x}) \mid x_1, \dots, x_{i-1}]
    \]
    \State Compute feature contribution:
    \[
    v(i, \mathbf{x}_{\text{test}}) \gets \mathbb{E}[f(\mathbf{x}) \mid x_1, \dots, x_i] - \mathbb{E}[f(\mathbf{x}) \mid x_1, \dots, x_{i-1}]
    \]
\EndFor

\State Verify local accuracy:
\[
f(\mathbf{x}_{\text{test}}) = v_0 + \sum_{i=1}^p v(i, \mathbf{x}_{\text{test}})
\]

\State Optional: Apply heuristic ordering of features based on marginal importance \(|\Delta_{j \mid \emptyset}(\mathbf{x}_{\text{test}})|\) to handle order sensitivity.

\State \textbf{return} $\{v(i, \mathbf{x}_{\text{test}})\}_{i=1}^p$
\end{algorithmic}
\end{algorithm}

By implementing the variable attribution method, as outlined in Algorithm~\ref{alg:variable_attribution}, the framework enables robust, feature-level explanations, empowering analysts to make data-driven decisions and refine IDS configurations effectively.

\subsection{Strategic Data Partitioning}

To stress-test the model’s generalization, we adopt a stringent partition:
\begin{itemize}
    \item \textbf{Training set:} $10\%$ of the available data.
    \item \textbf{Test set:} $90\%$ of the data.
\end{itemize}
Although unconventional, this setup closely mimics scenarios with limited labeled data, illuminating the framework’s capacity to operate under realistic constraints.

\subsection{Framework Contributions}

\textbf{Algorithm~\ref{alg:ids_framework}} outlines the complete pipeline of the proposed framework, integrating the components described above. The framework offers the following key contributions:

\begin{itemize}
    \item \textit{\textbf{Unified Preprocessing for Heterogeneous Datasets:}} Ensures consistency and scalability by standardizing input features across diverse datasets.
    \item \textit{\textbf{Latent Representation Learning with VAEs:}} Introduces dynamically calibrated VAEs to capture robust latent representations, enhancing anomaly detection.
    \item \textit{\textbf{Optimized Computational Efficiency:}} Leverages Knowledge Distillation to reduce model complexity, enabling deployment in resource-constrained environments.
    \item \textit{\textbf{Explainability with Variable Attribution:}} Employs a variable attribution-based approach for transparent decision-making, enhancing trust and usability for security analysts.
    \item \textit{\textbf{Rigorous Validation Across Scenarios:}} Implements a comprehensive evaluation strategy, showcasing robustness and adaptability under diverse operational constraints.
\end{itemize}

\begin{algorithm}[htbp] % Allows some flexibility in positioning
\caption{Proposed LENS-XAI Framework}
\label{alg:ids_framework}
\begin{algorithmic}[1]
\Require $\{D_i\}_{i=1}^M$: Set of $M$ raw datasets (e.g., CTU-13, UKM20, NSL-KDD, Edge-IIoTset)
\Ensure Anomaly predictions $\hat{y}$ with feature attributions $\{\phi_i\}$

\State \textbf{Initialization:}
\Statex \hspace{\algorithmicindent} Define Encoder/Decoder networks for VAE: $\{E,\,D\}$
\Statex \hspace{\algorithmicindent} Initialize Teacher and Student models: $\{f_t,\,f_s\}$

\State \textbf{Step 1: Unified Data Preprocessing}
\For{each dataset $D_i \in \{D_i\}$}
    \State Clean and preprocess:
    \Statex \hspace{\algorithmicindent} \(\bullet\) Handle missing values (imputation or removal)
    \Statex \hspace{\algorithmicindent} \(\bullet\) Encode categorical features (e.g., one-hot encoding)
    \Statex \hspace{\algorithmicindent} \(\bullet\) Normalize numerical features (e.g., Standard Scaler)
\EndFor
\State Merge preprocessed datasets into a unified dataset $D$

\State \textbf{Step 2: VAE Training for Latent Representations}
\For{each mini-batch $\mathbf{x}$ sampled from $D$}
    \State Compute latent embeddings: $\mathbf{z} \leftarrow E(\mathbf{x})$
    \State Reconstruct input: $\hat{\mathbf{x}} \leftarrow D(\mathbf{z})$
    \State Compute VAE loss:
    \[
    \mathcal{L}_{\text{VAE}} = \mathcal{L}_{\text{recon}} + \beta\,D_{\text{KL}}\big(q_\phi(\mathbf{z} \mid \mathbf{x}),\,p(\mathbf{z})\big)
    \]
    \State Update network parameters $\phi$ and $\theta$ via backpropagation
\EndFor
\State Record latent embeddings $\{\mathbf{z}\}$ after convergence

\State \textbf{Step 3: Knowledge Distillation}
\State Train teacher model $f_t$ on latent embeddings $\{\mathbf{z}\}$
\State Generate soft targets $\sigma_T(\hat{y}_t)$ using teacher model
\State Train student model $f_s$ to minimize the distillation loss:
\[
\mathcal{L}_{\text{distill}} = (1-\alpha)\,\mathcal{L}_{\text{CE}} + \alpha\,T^2\,D_{\text{KL}}\big(\sigma_T(\hat{y}_t), \,\sigma_T(\hat{y}_s)\big)
\]

\State \textbf{Step 4: Feature Attribution and Explainability}
\For{each test sample $\mathbf{x}_{\text{test}}$}
    \State Compute feature attributions $\phi_i(\mathbf{x}_{\text{test}})$ using variable attribution
    \State Rank or visualize features based on their contributions
\EndFor

\State \textbf{Step 5: Evaluation and Anomaly Detection}
\State Partition the data: \(10\%\) for training, \(90\%\) for testing
\State Use student model $f_s$ to predict anomaly scores $\hat{y}_s(\mathbf{x})$
\State Output anomaly scores and feature attributions $\{\phi_i\}$

\State \textbf{return} $\hat{y}, \{\phi_i\}$
\end{algorithmic}
\end{algorithm}

The integration of these elements positions the framework as a \textit{scalable} and \textit{interpretable} solution for intrusion detection, addressing challenges in modern IoT and edge computing environments.

\section{Performance evaluation}
\label{sec:4-evaluation}
We begin by assessing the intrusion detection performance of the proposed LENS-XAI framework, comparing it against baseline methods as outlined in Section~\ref{sec:baseline_methods}. Furthermore, Section~\ref{sec:explainability_evaluation} provides a comprehensive evaluation of the framework's explainability, benchmarked against state-of-the-art techniques.

\subsection{Experimental Setup and Evaluation Metrics}
The proposed framework is implemented in Python, incorporating modern libraries such as TensorFlow and PyTorch for model development and leveraging variable attribution methods for explainability. The preprocessing pipeline is designed to ensure data consistency by addressing missing values, encoding categorical variables, and normalizing numerical features. Explainability is achieved through an attribution-based approach that highlights the contribution of individual features to the model's predictions, enabling interpretable anomaly detection. The experimental settings use a batch size of 64 across all datasets. Training epochs are set to 200 for UKM-IDS20, 500 for Edge-IIoTset, and 100 for NSL-KDD, with a consistent learning rate of 0.001, ensuring robust latent representations. Knowledge distillation is conducted with a temperature parameter \(T = 2\) and a weighting coefficient \(\alpha = 0.5\). The system configuration, tailored to support the integration of XAI principles, is summarized in Table~\ref{tab:configurations}.

\begin{table*}[h!]
\centering
\caption{Software and hardware configurations for experimentation.}
\vspace{0.5em}
\label{tab:configurations}
\resizebox{\textwidth}{!}{%
\begin{tabular}{lp{10cm}}
\toprule
\hline
\textbf{Specification} & \textbf{Details} \\ 
\hline
\textsc{Operating System} & Windows 11 \\ 
\hline
\textsc{Python Environment} & Anaconda Navigator, Python 3.11 \\ 
\hline
\textsc{Libraries} & TensorFlow, PyTorch, Dalex \\ 
\hline
\textsc{Processor} & Intel Core i7-14700K @ 3.4 GHz \\ 
\hline
\textsc{RAM} & 32 GB DDR5 \\ 
\hline
\textsc{Storage} & 2 TB SSD \\ 
\hline
\textsc{GPU} & NVIDIA GeForce RTX 4080 Super (16 GB VRAM) \\ 
\hline
\bottomrule
\end{tabular}%
}
\end{table*}

The key evaluation metrics used in this study are defined as follows:

\begin{itemize}
    \item \textbf{Performance Metrics}: The model's detection performance was evaluated using metrics such as accuracy, precision, recall, and  \(F_1\)-score. These metrics provided a comprehensive understanding of the model's ability to detect various types of cyberattacks and minimize errors.

    \item \textbf{Number of Parameters}: The total number of learnable parameters in the model was calculated as:
\begin{equation}
    \text{Number of Parameters} = \sum_{i=1}^{L} \text{Parameters}(i),
\end{equation}
    where \(L\) is the number of layers in the model, and \(\textit{Parameters}(i)\) represents the parameters in the \(i\)-th layer.

    \item \textbf{Inference Time per Batch}: The time required for the model to process a batch of input data was calculated as:
\begin{equation}
    \text{Inference Time (ms)} = \frac{\text{End Time} - \text{Start Time}}{\text{Batch Size}}.
\end{equation}

    \item \textbf{Memory Usage}: The memory footprint of the model during inference was evaluated to determine the hardware requirements for real-time deployment.
\end{itemize}

\subsection{Dataset Descriptions}
%\label{ss:datasets}

The proposed LENS-XAI framework is assessed using four benchmark datasets: \textit{Edge-IIoTset, UKM20, CTU-13, and NSL-KDD}. These datasets were carefully chosen for their distinctive features and their relevance to intrusion detection in IIoT and Vehicular Network (IVN) environments. Table~\ref{tab:dataset_comparison}, detailed in Appendix I, summarizes the key characteristics of these datasets, highlighting their relevance for evaluating the proposed intrusion detection framework.

\begin{table}[h!]
\centering
\caption{Comparison of Edge-IIoTset, UKM20, CTU-13, and NSL-KDD datasets.}
\label{tab:dataset_comparison}
\vspace{0.5em}
\resizebox{\textwidth}{!}{%
\begin{tabular}{p{2.8cm}p{3.8cm}p{3.8cm}p{3.8cm}p{3cm}}
\toprule
\hline
\textbf{Dataset} & \textbf{Benefits} & \textbf{Challenges} & \textbf{Attacks Included} & \textbf{Number of Samples} \\ 
\midrule
\hline
\textsc{Edge-IIoTset} & 
Comprehensive IoT/IIoT coverage; supports federated learning; diverse attack scenarios. & 
Data complexity; preprocessing required for analysis. & 
Backdoor, DDoS HTTP, DDoS ICMP, DDoS TCP, DDoS UDP, Fingerprinting, MITM, Password, Port Scanning, Ransomware, SQL Injection, Uploading, Vulnerability Scanner, XSS & 
61 of 1176 features; 2,219,201 samples \\ 
\hline
\textsc{UKM20} & 
Focuses on UAV communication security; high accuracy; includes diverse attacks. & 
Relatively new dataset; requires validation and acceptance. & 
ARP Poisoning, BeEF HTTP Exploits, Mass HTTP Requests, Metasploit Exploits, Port Scanning, TCP Flood, UDP Data Flood & 
46 UAV-related features; 12,887 samples \\ 
\hline
\textsc{CTU-13} & 
Real botnet traffic mixed with normal traffic; widely used. & 
Imbalanced attack classes; limited scenarios. & 
Botnet, DDoS, Port Scans & 
Bidirectional NetFlows; 92,212 samples \\ 
\hline
\textsc{NSL-KDD} & 
Enhanced KDD Cup 99; reduced redundancy and imbalance. & 
Class imbalance persists; limited modern threat coverage. & 
DoS, Probe, R2L, U2R & 
41 features; 148,517 samples \\ 
\bottomrule
\hline
\end{tabular}%
}
\end{table}

\begin{comment}

\begin{table}[h]
    \centering
    \scriptsize
    \renewcommand{\arraystretch}{1.5} % Increase row height
    \caption{Comparison of Benchmark Datasets for Intrusion Detection Systems}
    \label{tab:dataset_comparison}
    \begin{tabular}{|p{2.5cm}|p{4.5cm}|p{3cm}|p{2.5cm}|}
        \hline
        \textbf{Dataset} & \textbf{Key Features} & \textbf{Attack Types} & \textbf{Sample Count} \\ 
        \hline
        \textsc{Edge-IIoTset} & Comprehensive IoT/IIoT coverage; supports federated learning; diverse attack scenarios & DoS, DDoS, Man-in-the-Middle, Injection, Malware & 2,219,201 \\ 
        \hline
        \textsc{UKM-IDS20} & Focused on UAV security; high detection accuracy; diverse attack types & ARP poisoning, DoS, Scans, Exploits & 12,887 \\ 
        \hline
        \textsc{CTU-13} & Real botnet traffic with normal traffic; widely used in research & Botnet attacks (various types), DDoS, Port Scans & 92,212 \\ 
        \hline
        \textsc{NSL-KDD} & Enhanced KDD Cup 99; addresses redundancy and imbalance issues & DoS, Probe, R2L (Remote to Local), U2R (User to Root) & 148,517 \\ 
        \hline
    \end{tabular}
\end{table}

\end{comment}

\begin{figure}[ht]
    \centering
    \includegraphics[width=0.8\textwidth]{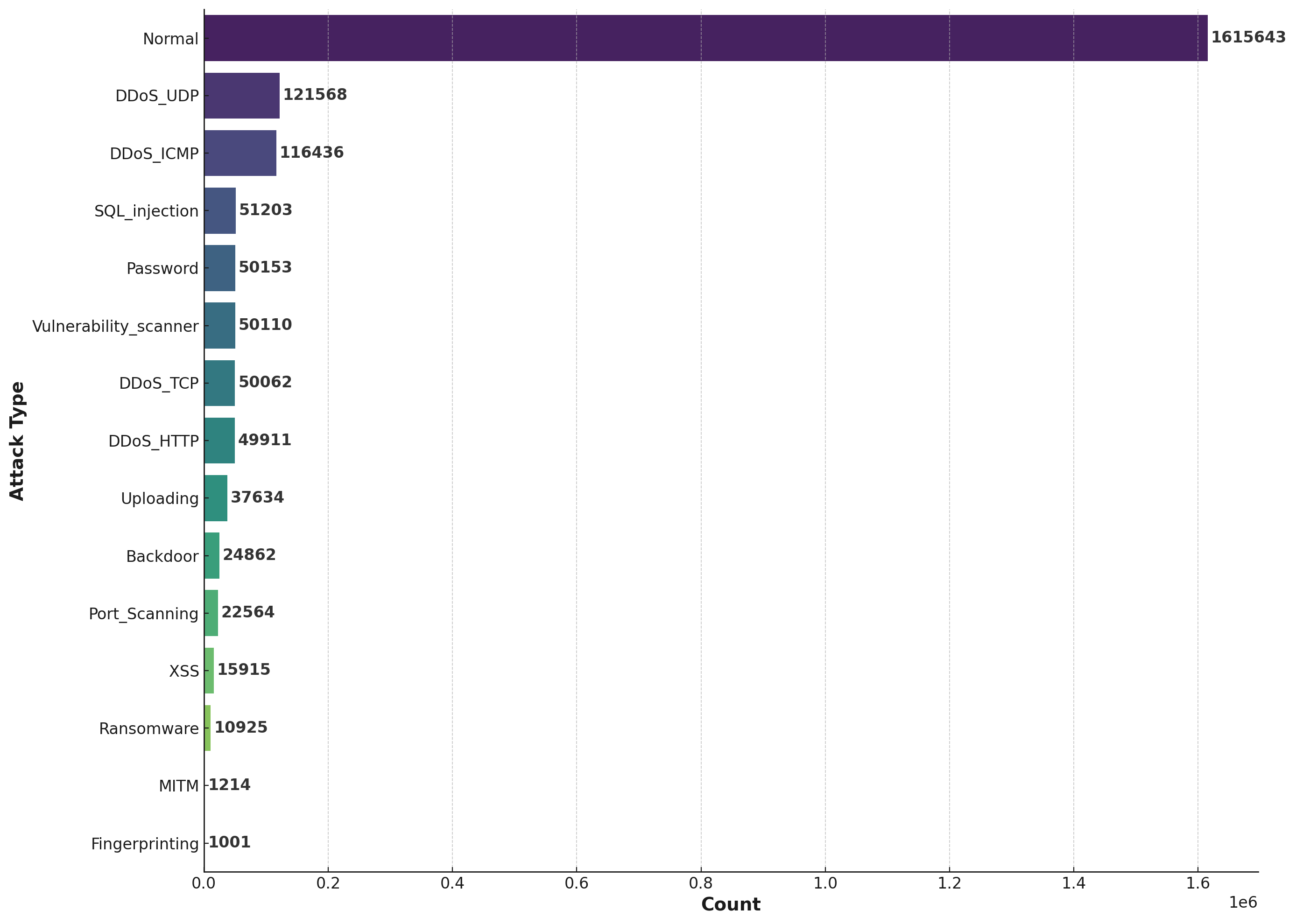}
    \caption{Class distribution of Edge-IIoTset dataset.}
    \label{fig:edgeiot_class}
\end{figure}

\paragraph{\textbf{Edge-IIoTset Dataset}}
The Edge-IIoTset dataset is designed for cybersecurity research in IoT and edge computing environments. It features both benign and malicious network traffic, annotated with detailed labels, timestamps, and traffic characteristics. The dataset spans multiple protocols and includes diverse attack types such as Denial of Service (DoS), data exfiltration, and command injection. This comprehensive coverage makes Edge-IIoTset an invaluable resource for testing intrusion detection systems in complex and realistic network scenarios. Figure~\ref{fig:edgeiot_class} shows the various classes of the Edge-IIoTset. This dataset is publicly accessible at\footnote{\fontsize{6}{6}\selectfont\url{https://ieee-dataport.org/documents/edge-iiotset-new-comprehensive-realistic-cyber-security-dataset-iot-and-iiot-applications}}.

\begin{figure}[ht]
    \centering
    \includegraphics[width=0.8\textwidth]{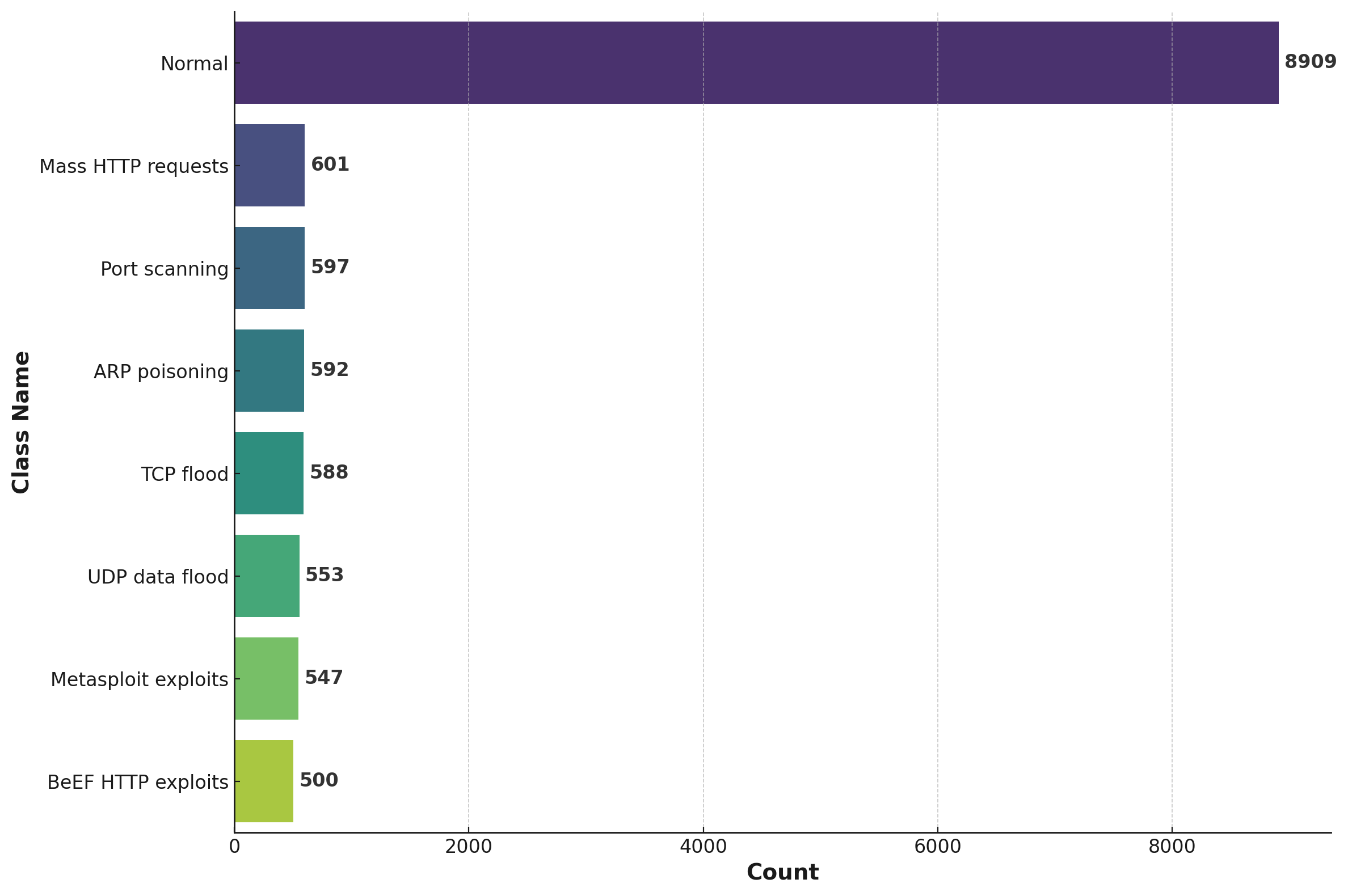}
    \caption{Class distribution of UKM20 dataset.}
    \label{fig:UKM20_class}
\end{figure}

\paragraph{\textbf{UKM20 Dataset}}  
The UKM-IDS20 dataset is tailored for intrusion detection in vehicular networks, offering traffic data from Controller Area Network (CAN) buses. It encompasses both normal and attack scenarios, including message injection and spoofing attacks. Key features, such as CAN IDs, data length codes (DLC), and payload data, enable robust anomaly detection and the evaluation of IDS solutions specific to IVN environments. Figure~\ref{fig:UKM20_class} shows the class distribution of the UKM20. The dataset is publicly accessible at\footnote{\fontsize{7}{7}\selectfont\url{https://www.kaggle.com/datasets/muatazsalam/ukm-ids20}}.

\paragraph{\textbf{CTU-13 Dataset}}  
The CTU-13 dataset is a widely recognized benchmark for botnet traffic analysis. While not specific to IVNs, it contains a mix of normal and botnet-infected traffic with detailed flow-based features. The dataset is well-suited for evaluating generalized IDS models that can be adapted for IVN use cases. The CTU-13 dataset is publicly available at\footnote{\fontsize{7}{7}\selectfont\url{https://www.stratosphereips.org/datasets-ctu13}}. This dataset consists of the following traffic types and their respective counts:
\begin{itemize}
    \item \textbf{Normal Traffic:} 53,314 instances,
    \item \textbf{Attack Traffic:} 38,898 instances.
\end{itemize}

\begin{figure}[ht]
    \centering
    \includegraphics[width=0.8\textwidth]{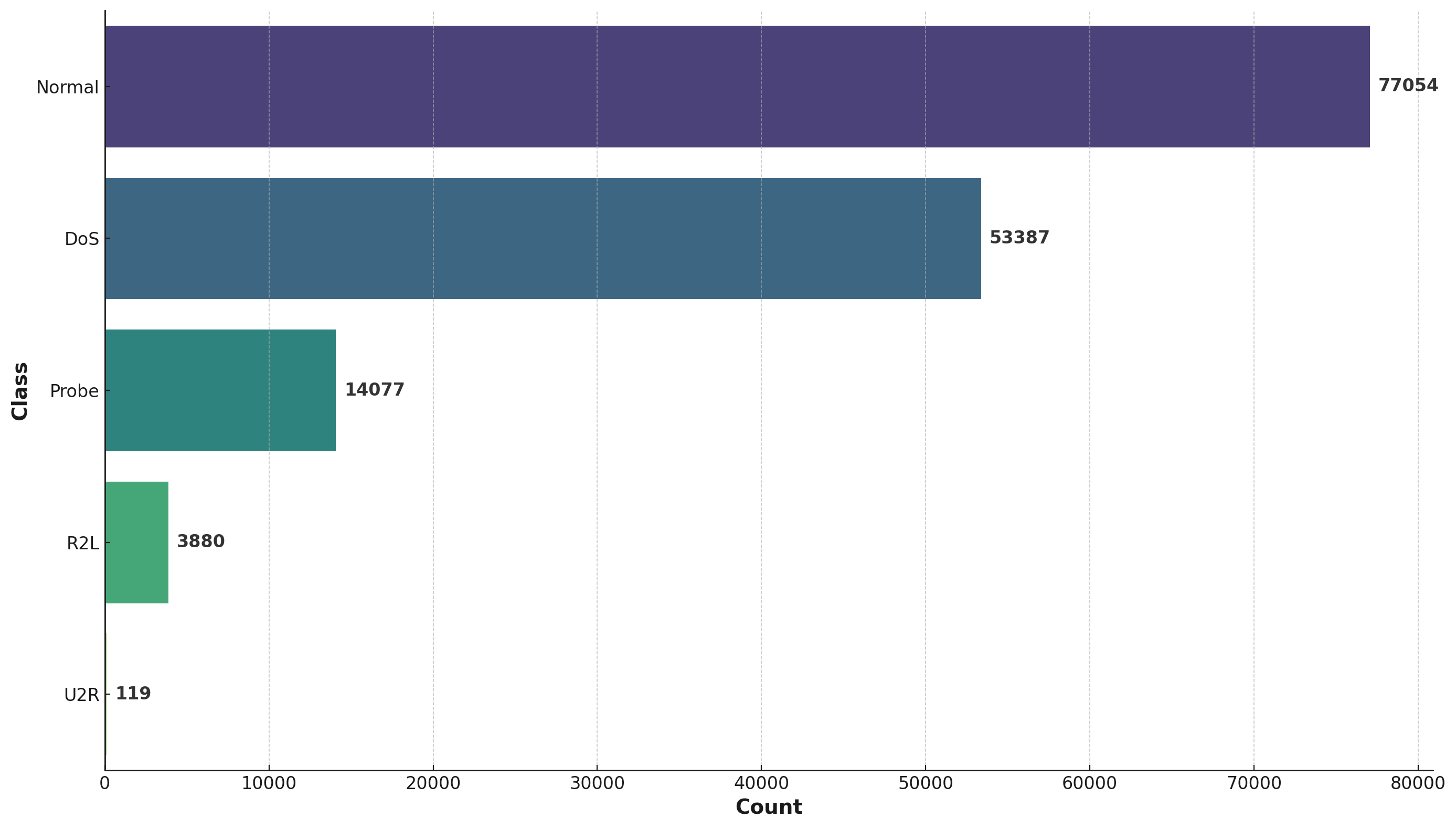}
    \caption{Class distribution of NSL-KDD dataset.}
    \label{fig:NSL-KDD_class}
\end{figure}

\paragraph{\textbf{NSL-KDD Dataset}}  
The NSL-KDD dataset is a refined version of the KDD Cup 1999 dataset, addressing redundancy and imbalance issues. It provides labeled network traffic with 43 features, including protocol type, service, flag, and attack labels. The dataset includes separate training (KDDTrain+) and testing (KDDTest+) subsets, facilitating model development and validation in diverse cybersecurity contexts. Figure~\ref{fig:NSL-KDD_class} illustrates the distribution of classes within the NSL-KDD. This dataset is publicly available at\footnote{\fontsize{7}{7}\selectfont\url{https://www.kaggle.com/datasets/hassan06/nslkdd}}.

\subsection{Performance Evaluation of the Introduced LENS-XAI Method}
\label{sec:baseline_methods}

\subsubsection{Analysis on Edge-IIoTset Dataset}
Table~\ref{tab:edge_iiotset_comparisonn} provides a comprehensive comparison of the proposed \textit{LENS-XAI} framework with state-of-the-art intrusion detection systems on the Edge-IIoTset dataset, evaluating multi-class classification tasks.

\paragraph{\textbf{Multi-class Classification Results}}
\begin{itemize}
    \item \textbf{Accuracy}: The \textit{LENS-XAI Teacher} model achieved the highest accuracy of \textbf{95.34\%}, slightly outperforming the \textit{Student} model at \textbf{95.31\%}, as well as other models such as NIDS-BAI (\textbf{94.7\%}) and BGA (\textbf{94.2\%}) \cite{yang2024improved}. These results highlight the robustness of the framework in handling complex IoT-specific attack scenarios.

    \item \textbf{Precision and Recall}: The \textit{Teacher} model demonstrated superior precision (\textbf{96.75\%}) and recall (\textbf{95.34\%}) compared to NIDS-BAI (\textbf{94.7\% precision}, \textbf{94.8\% recall}) \cite{yang2024improved}. Similarly, the \textit{Student} model maintained competitive precision (\textbf{95.74\%}) and recall (\textbf{95.31\%}), showcasing effective knowledge transfer and computational efficiency.

    \item \textbf{F1-Measure}: Both models achieved balanced performance, with the \textit{Teacher} model scoring an F1-measure of \textbf{95.09\%} and the \textit{Student} model closely following with \textbf{95.36\%}. These values surpass traditional models like ICNN (\textbf{91.6\%}) and BGA (\textbf{94.3\%}) \cite{yang2024improved}, underlining the versatility of \textit{LENS-XAI} in IoT-based security tasks.
\end{itemize}

\begin{table}[h!]
\centering
\caption{Performance metrics comparison for multi-class classification on the Edge-IIoTset dataset.}
\vspace{0.5em}
\resizebox{\textwidth}{!}{%
\label{tab:edge_iiotset_comparisonn}
\begin{tabular}{@{}lcccc@{}}
\toprule
\hline
\textbf{Model} & \textbf{Accuracy (\%)} & \textbf{Precision (\%)} & \textbf{Recall (\%)} & \textbf{F1-Score (\%)} \\ 
\midrule
\hline
BGA \cite{yang2024improved} & 94.20 & 94.20 & 94.40 & 94.30 \\
ICNN \cite{yang2024improved} & 91.80 & 91.80 & 94.20 & 91.60 \\
NIDS-BAI \cite{yang2024improved} & 94.70 & 94.70 & 94.80 & 94.60 \\
\rowcolor{green!20} \textbf{LENS-XAI Teacher} & \textbf{95.34} & \textbf{96.75} & \textbf{95.34} & \textbf{95.09} \\
\rowcolor{green!20} \textbf{LENS-XAI Student} & \textbf{95.31} & \textbf{95.74} & \textbf{95.31} & \textbf{95.36} \\
\hline
\bottomrule
\end{tabular}
}
\end{table}

\paragraph{\textbf{Performance Analysis Across Attack Types}}
Table~\ref{tab:eval_hcrl} highlights the LENS-XAI framework’s performance across various attack types for the Edge-IIoTset dataset. Both the \textit{Teacher} and \textit{Student} models excel in detecting frequent attacks like \textbf{DDoS ICMP} and \textbf{DDoS UDP}, achieving near-perfect accuracy (\textbf{99.64\%}–\textbf{100.00\%}). For \textbf{Backdoor}, the models perform strongly (\textbf{97.89\%}–\textbf{98.49\%}), while in rare attacks like \textbf{SQL Injection}, the \textit{Teacher} outperforms the \textit{Student} (\textbf{34.21\%} vs. \textbf{19.13\%}). Both models achieve perfect detection (\textbf{100.00\%}) for \textbf{Normal} traffic, showcasing reliability. The \textit{Teacher} demonstrates better handling of rare scenarios, solidifying its robustness for IoT intrusion detection.

\begin{table*}[t]
\centering 
\caption{Performance metrics comparison for the LENS-XAI framework on the Edge-IIoTset dataset across various attack types.}
\label{tab:eval_hcrl}
\vspace{0.5em}
\begin{small}
\resizebox{\textwidth}{!}{%
\begin{tabular}{lcccc}
\toprule
\hline

\textbf{Attack Type} & \textbf{Accuracy (\%)} & \textbf{Recall (\%)} & \textbf{Precision (\%)} & \textbf{F1 Score (\%)} \\ 
\midrule
\hline

\rowcolor{green!20}
\textbf{Backdoor (Teacher)} & \textbf{97.89} & \textbf{97.89} & \textbf{99.18} & \textbf{98.53} \\ 
\rowcolor{green!20}
\textbf{Backdoor (Student)} & \textbf{98.49} & \textbf{98.49} & \textbf{97.99} & \textbf{98.24} \\ 
\hline

DDoS HTTP (Teacher) & 82.81 & 82.81 & 94.58 & 88.30 \\ 
DDoS HTTP (Student) & 83.07 & 83.07 & 94.41 & 88.38 \\ 
\hline

\rowcolor{green!20}
\textbf{DDoS ICMP (Teacher)} & \textbf{99.64} & \textbf{99.64} & \textbf{100.00} & \textbf{99.82} \\ 
\rowcolor{green!20}
\textbf{DDoS ICMP (Student)} & \textbf{99.90} & \textbf{99.90} & \textbf{99.99} & \textbf{99.95} \\ 
\hline

\rowcolor{green!20}
\textbf{DDoS TCP (Teacher)} & \textbf{99.74} & \textbf{99.74} & \textbf{85.33} & \textbf{91.98} \\ 
\rowcolor{green!20}
\textbf{DDoS TCP (Student)} & \textbf{99.03} & \textbf{99.03} & \textbf{76.02} & \textbf{86.01} \\ 
\hline

\rowcolor{green!20}
\textbf{DDoS UDP (Teacher)} & \textbf{100.00} & \textbf{100.00} & \textbf{99.75} & \textbf{99.87} \\ 
\rowcolor{green!20}
\textbf{DDoS UDP (Student)} & \textbf{100.00} & \textbf{100.00} & \textbf{99.87} & \textbf{99.93} \\ 
\hline

Fingerprinting (Teacher) & 60.66 & 60.66 & 100.00 & 75.51 \\ 
Fingerprinting (Student) & 57.63 & 57.63 & 100.00 & 73.12 \\ 
\hline

\rowcolor{green!20}
\textbf{MITM (Teacher)} & \textbf{93.01} & \textbf{93.01} & \textbf{100.00} & \textbf{96.38} \\ 
\rowcolor{green!20}
\textbf{MITM (Student)} & \textbf{93.01} & \textbf{93.01} & \textbf{100.00} & \textbf{96.38} \\ 
\hline

\rowcolor{green!20}
\textbf{Normal (Teacher)} & \textbf{100.00} & \textbf{100.00} & \textbf{100.00} & \textbf{100.00} \\ 
\rowcolor{green!20}
\textbf{Normal (Student)} & \textbf{100.00} & \textbf{100.00} & \textbf{100.00} & \textbf{100.00} \\ 
\hline

Password (Teacher) & 70.98 & 70.98 & 45.58 & 55.51 \\ 
Password (Student) & 97.28 & 97.28 & 42.33 & 58.99 \\ 
\hline

Port Scanning (Teacher) & 58.07 & 58.07 & 94.86 & 72.04 \\ 
Port Scanning (Student) & 21.35 & 21.35 & 90.01 & 34.51 \\ 
\hline

\rowcolor{green!20}
\textbf{Ransomware (Teacher)} & \textbf{96.14} & \textbf{96.14} & \textbf{99.92} & \textbf{97.99} \\ 
\rowcolor{green!20}
\textbf{Ransomware (Student)} & \textbf{96.62} & \textbf{96.62} & \textbf{97.06} & \textbf{96.84} \\ 
\hline

SQL Injection (Teacher) & 34.21 & 34.21 & 58.70 & 43.23 \\ 
SQL Injection (Student) & 19.13 & 19.13 & 64.96 & 29.55 \\ 
\hline

Uploading (Teacher) & 56.10 & 56.10 & 63.60 & 59.62 \\ 
Uploading (Student) & 27.28 & 27.28 & 99.72 & 42.84 \\ 
\hline

\rowcolor{green!20}
\textbf{Vulnerability Scanner (Teacher)} & \textbf{95.28} & \textbf{95.28} & \textbf{76.86} & \textbf{85.09} \\ 
\rowcolor{green!20}
\textbf{Vulnerability Scanner (Student)} & \textbf{84.85} & \textbf{84.85} & \textbf{94.72} & \textbf{89.51} \\ 
\hline

XSS (Teacher) & 30.36 & 30.36 & 62.70 & 40.91 \\ 
XSS (Student) & 77.66 & 77.66 & 47.94 & 59.28 \\ 
\bottomrule
\hline
\end{tabular}%
}
\end{small}
\end{table*}

\begin{figure}[ht]
    \centering
    \includegraphics[width=\textwidth]{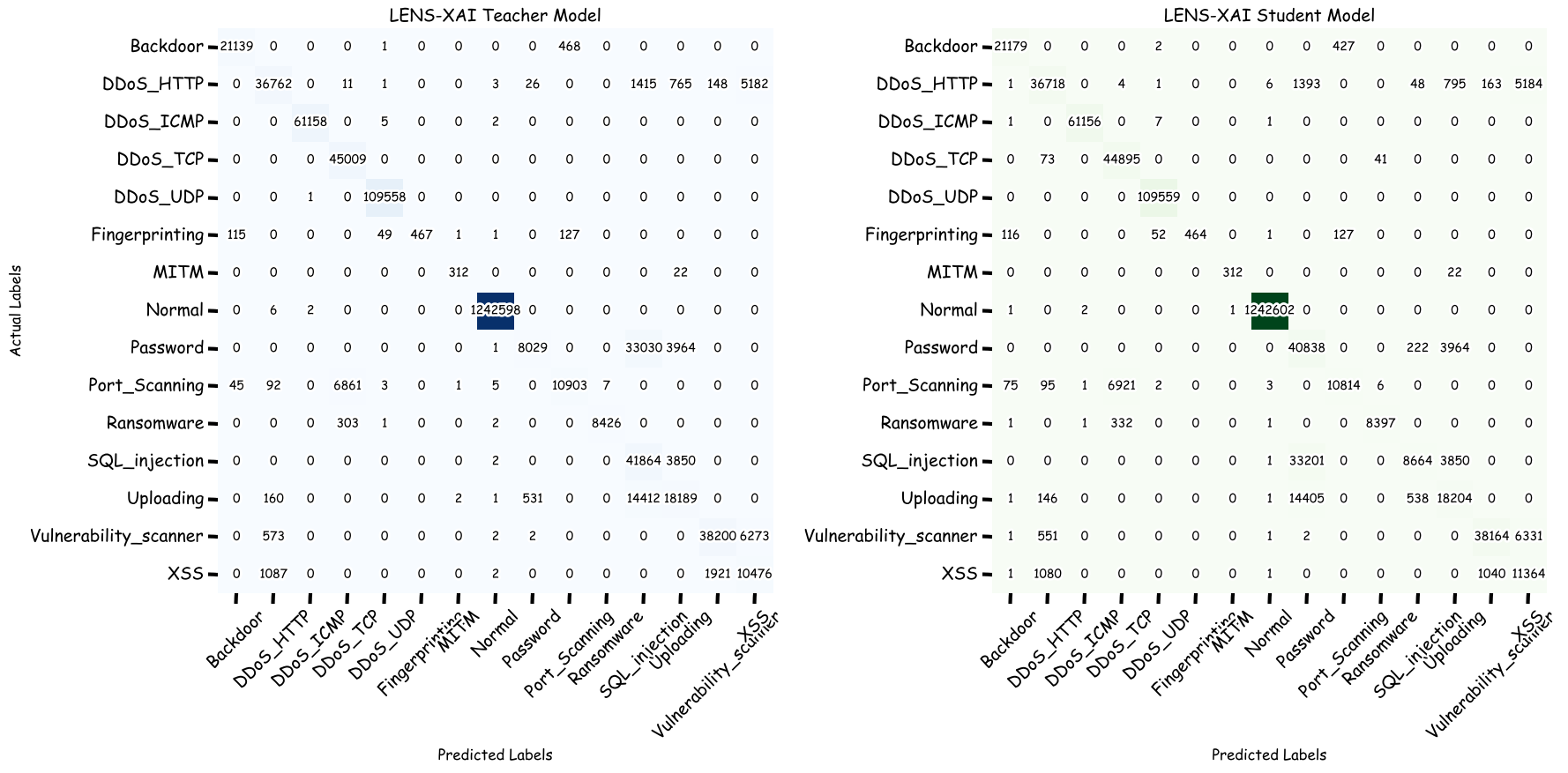}
    \vspace{-0,5em}
    \caption{Confusion matrices of the LENS-XAI framework on the Edge-IIoTSet dataset: \textit{(Left)} Teacher  and \textit{(Right)} Student models, showcasing effective differentiation between 15 classes.}
    \label{fig:Edge-IIoTSetMulticlass_confusion}
\end{figure}

\paragraph{\textbf{Confusion Matrix Analysis}} Figure~\ref{fig:Edge-IIoTSetMulticlass_confusion} illustrates the confusion matrices for the LENS-XAI framework, showcasing the performance of the \textit{Teacher} and \textit{Student} models on the Edge IIoT dataset for multiclass classification. The \textit{Teacher} model achieves high accuracy across frequent attacks such as \textbf{DDoS ICMP} and \textbf{DDoS UDP}, with minimal misclassifications. For instance, only 1 instance of \textbf{DDoS ICMP} was misclassified as another type, demonstrating robust detection capabilities.

Similarly, the \textit{Student} model exhibits strong performance, closely approximating the \textit{Teacher}'s accuracy. Notably, it correctly classifies 61,156 \textbf{DDoS ICMP} instances and 109,559 \textbf{DDoS UDP} instances. While both models perform exceptionally well for frequent attack types, rare attack types like \textbf{SQL Injection} and \textbf{Uploading} exhibit slightly higher misclassification rates. These results affirm the scalability and reliability of the LENS-XAI framework in detecting a wide range of attack types within IoT environments.

\subsubsection{Analysis on UKM20 Dataset}
Table~\ref{tab:ukm20_multiclasss} provides a comparison of the performance metrics for the \textit{LENS-XAI} framework, highlighting its effectiveness in multi-class classification tasks on the UKM20 dataset.
\vspace{2em}
\paragraph{\textbf{Multi-class Classification Results}}
\begin{itemize}
    \item \textbf{Accuracy}: The \textit{LENS-XAI Teacher} model achieved a near-perfect accuracy of \textbf{99.92\%}, outperforming the \textit{Student} model's accuracy of \textbf{99.80\%}. These results demonstrate the robustness of the \textit{Teacher} model in maintaining classification consistency, with the \textit{Student} model closely approximating this performance.

    \item \textbf{Precision and Recall}: Both models exhibit exceptionally high precision and recall values, with the \textit{Teacher} model achieving \textbf{99.92\%} for both metrics, slightly higher than the \textit{Student} model’s \textbf{99.80\%}. These values highlight the capability of the \textit{LENS-XAI} framework to accurately identify attack types while minimizing false positives and false negatives.

    \item \textbf{F1-Measure}: The \textit{Teacher} model demonstrated an F1-measure of \textbf{99.92\%}, indicating a balanced performance across precision and recall. The \textit{Student} model also performed remarkably, achieving an F1-measure of \textbf{99.80\%}, validating the efficacy of knowledge distillation in the \textit{LENS-XAI} framework.
\end{itemize}

\begin{table}[h!]
\centering
\caption{Evaluation metrics for multi-class classification on the UKM20 dataset.}
\vspace{0.5em}
\resizebox{0.95\textwidth}{!}{%
\label{tab:ukm20_multiclasss}
\begin{tabular}{@{}lcccc@{}}
\toprule
\hline
\textbf{Model} & \textbf{Accuracy (\%)} & \textbf{Precision (\%)} & \textbf{Recall (\%)} & \textbf{F1-Score (\%)} \\ 
\midrule
\hline
\rowcolor{green!20} \textbf{LENS-XAI Teacher} & \textbf{99.92} & \textbf{99.92} & \textbf{99.92} & \textbf{99.92} \\
\rowcolor{green!20} \textbf{LENS-XAI Student} & \textbf{99.80} & \textbf{99.80} & \textbf{99.80} & \textbf{99.80} \\
\hline
\bottomrule
\end{tabular}
}
\end{table}

\paragraph{\textbf{Performance Analysis Across Attack Types}}
Table~\ref{tab:ukm_model_comparisonn} highlights the performance of the \textit{LENS-XAI Teacher} and \textit{Student} models across various attack types in the UKM20 dataset. 

\begin{itemize}
    \item \textbf{Frequent Attacks}: Both models excel in detecting common attacks such as \textbf{BeEF HTTP Exploits}, \textbf{Metasploit Exploits}, and \textbf{Normal} traffic, achieving perfect or near-perfect accuracy (\textbf{99.95\%}–\textbf{100.00\%}). The \textit{Teacher} and \textit{Student} models are equally capable in these scenarios, demonstrating high reliability for frequent attack detection.

    \item \textbf{Rare Attacks}: For less frequent attacks like \textbf{ARP Poisoning} and \textbf{Mass HTTP Requests}, the \textit{Student} model slightly outperforms the \textit{Teacher} model in accuracy (\textbf{99.77\%} vs. \textbf{95.72\%} for ARP Poisoning). This reflects the effectiveness of knowledge distillation in the \textit{LENS-XAI} framework, enabling the \textit{Student} model to closely replicate the \textit{Teacher} model's performance.
\end{itemize}

\begin{table*}[t]
\centering 
\caption{
Performance metrics comparison for the LENS-XAI framework on the UKM20 dataset across various attack types.}
\label{tab:ukm_model_comparisonn}
\vspace{0.5em}
\begin{small}
\resizebox{\textwidth}{!}{%
\begin{tabular}{lcccc}
\toprule
\hline
\textbf{Attack Type} & \textbf{Accuracy (\%)} & \textbf{Precision (\%)} & \textbf{Recall (\%)} & \textbf{F1 Score (\%)} \\ 
\midrule

\rowcolor{green!20}
ARP Poisoning (Teacher) & \textbf{95.72} & \textbf{82.05} & \textbf{95.72} & \textbf{88.36} \\ 
\rowcolor{green!20}
ARP Poisoning (Student) & \textbf{99.77} & \textbf{81.89} & \textbf{99.77} & \textbf{89.95} \\ 
\hline

\rowcolor{green!20}
BeEF HTTP Exploits (Teacher) & \textbf{100.00} & \textbf{100.00} & \textbf{100.00} & \textbf{100.00} \\ 
\rowcolor{green!20}
BeEF HTTP Exploits (Student) & \textbf{100.00} & \textbf{99.55} & \textbf{100.00} & \textbf{99.77} \\ 
\hline

\rowcolor{green!20}
Mass HTTP Requests (Teacher) & \textbf{99.26} & \textbf{99.81} & \textbf{99.26} & \textbf{99.53} \\ 
\rowcolor{green!20}
Mass HTTP Requests (Student) & \textbf{98.89} & \textbf{100.00} & \textbf{98.89} & \textbf{99.44} \\ 
\hline

\rowcolor{green!20}
Metasploit Exploits (Teacher) & \textbf{100.00} & \textbf{99.39} & \textbf{100.00} & \textbf{99.69} \\ 
\rowcolor{green!20}
Metasploit Exploits (Student) & \textbf{100.00} & \textbf{99.39} & \textbf{100.00} & \textbf{99.69} \\ 
\hline

\rowcolor{green!20}
Normal (Teacher) & \textbf{99.95} & \textbf{99.99} & \textbf{99.95} & \textbf{99.97} \\ 
\rowcolor{green!20}
Normal (Student) & \textbf{99.94} & \textbf{99.95} & \textbf{99.94} & \textbf{99.94} \\ 
\hline

\rowcolor{green!20}
Port Scanning (Teacher) & \textbf{99.81} & \textbf{100.00} & \textbf{99.81} & \textbf{99.91} \\ 
\rowcolor{green!20}
Port Scanning (Student) & \textbf{99.43} & \textbf{100.00} & \textbf{99.43} & \textbf{99.72} \\ 
\hline

\rowcolor{green!20}
TCP Flood (Teacher) & \textbf{100.00} & \textbf{99.44} & \textbf{100.00} & \textbf{99.72} \\ 
\rowcolor{green!20}
TCP Flood (Student) & \textbf{100.00} & \textbf{99.62} & \textbf{100.00} & \textbf{99.81} \\ 
\hline

\rowcolor{green!20}
UDP Data Flood (Teacher) & \textbf{100.00} & \textbf{99.80} & \textbf{100.00} & \textbf{99.90} \\ 
\rowcolor{green!20}
UDP Data Flood (Student) & \textbf{100.00} & \textbf{99.41} & \textbf{100.00} & \textbf{99.71} \\ 
\hline

\end{tabular}%
}
\end{small}
\end{table*}

\begin{figure}[ht]
    \centering
    \includegraphics[width=\textwidth]{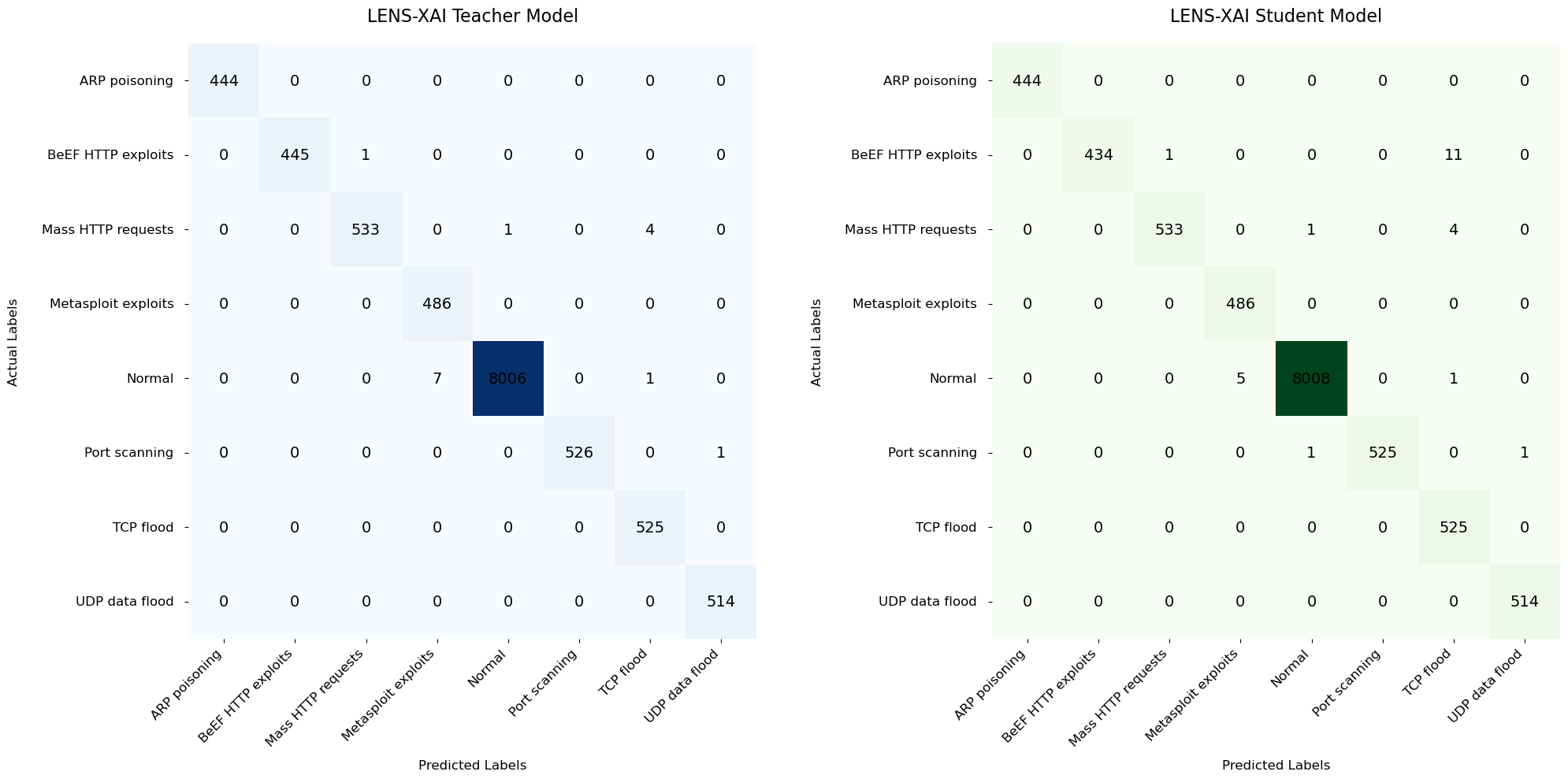}
    \vspace{-0.5em}
    \caption{Confusion matrices of the LENS-XAI framework on the UKM IDS20 dataset: \textit{(Left)} Teacher  and \textit{(Right)} Student models, showcasing effective differentiation between eight classes.}
    \label{fig:UKMMulticlass_confusion}
\end{figure}

\paragraph{\textbf{Confusion Matrix Analysis}}
Figure~\ref{fig:UKMMulticlass_confusion} presents the confusion matrices for the \textit{LENS-XAI} framework, illustrating the performance of the \textit{Teacher} and \textit{Student} models on the UKM-IDS20 dataset for multiclass classification tasks. The \textit{Teacher} model demonstrates high accuracy across all classes, with minimal misclassifications. For example, all 444 instances of \textbf{ARP Poisoning} are correctly classified, while only 7 \textbf{Normal} instances are misclassified. Similarly, frequent attack types such as \textbf{BeEF HTTP Exploits}, \textbf{Mass HTTP Requests}, and \textbf{Metasploit Exploits} achieve near-perfect detection, highlighting the model's robustness.

The \textit{Student} model closely approximates the \textit{Teacher}'s performance, with slightly higher misclassifications in certain categories. For instance, it correctly classifies 434 out of 445 \textbf{BeEF HTTP Exploits} instances, while 11 instances are misclassified. Despite these minor differences, the \textit{Student} model maintains strong overall performance, correctly identifying 8,008 \textbf{Normal} instances and 525 \textbf{TCP Flood} instances.

\subsubsection{Analysis on CTU-13 Dataset}
Table~\ref{tab:ml_models} presents a comparative analysis of the proposed LENS-XAI framework against state-of-the-art ML models on the CTU-13 dataset. The results underline the robustness and effectiveness of the proposed framework in intrusion detection.

\begin{itemize}
    \item \textbf{Accuracy}: The \textit{LENSXAI Teacher} achieved an exceptional accuracy of \textbf{98.42\%}, outperforming models like SVM-RBF (85.03\%) and k-NN (93.00\%) \cite{ibrahim2021multilayer}. The \textit{LENSXAI Student} also delivered a strong performance with an accuracy of \textbf{98.12\%}, demonstrating the efficiency of the knowledge distillation process.

   \item \textbf{Recall}: Both \textit{LENSXAI Teacher} and \textit{Student} achieved outstanding recall scores of \textbf{98.42\%} and \textbf{98.12\%}, respectively, ensuring high detection rates for malicious activities. These results outperform traditional methods such as NB (76.00\%) and LR (86.00\%) \cite{10593100}.

\item \textbf{Precision}: The precision scores of \textbf{98.43\%} for the \textit{Teacher} model and \textbf{98.14\%} for the \textit{Student} model highlight their capability to minimize false positives, ensuring reliability in real-world applications.

\item \textbf{F1 Score and Balanced Performance}: The harmonic mean of precision and recall for \textit{LENSXAI Teacher} and \textit{Student} were \textbf{98.42\%} and \textbf{98.12\%}, respectively, showcasing balanced and consistent performance in intrusion detection tasks.

\item \textbf{Comparison with State-of-the-Art}:
\begin{itemize}
    \item \textbf{FFS-HTTP \cite{letteri2019feature}}: Although achieving slightly higher accuracy (98.79\%), FFS-HTTP lacks the interpretability and lightweight design offered by LENS-XAI.
    \item \textbf{bot-DL \cite{10.1002/nem.2039}}: Demonstrated competitive accuracy (96.60\%) but lacks the explainability critical for trust and transparency in intrusion detection systems, a key advantage of LENS-XAI.
    \item \textbf{k-NN and SVM-RBF \cite{ibrahim2021multilayer}}: Fell short in both accuracy and scalability, with lower accuracy scores of 93.00\% and 85.03\%, respectively, underlining the efficacy of the advanced techniques integrated into LENS-XAI.
\end{itemize}

\end{itemize}

\begin{table}[h!]
\centering
\caption{Classification achievements of the proposed scheme on the CTU-13 dataset.}
\vspace{0.5em}
\resizebox{0.97\textwidth}{!}{%
\label{tab:ml_models}
\begin{tabular}{@{}lcccc@{}}
\toprule
\hline
\textbf{Model} & \textbf{Accuracy (\%)} & \textbf{Precision (\%)} & \textbf{Recall (\%)} & \textbf{F1-Score (\%)} \\
\midrule
\hline
FFS-HTTP \cite{letteri2019feature} & 98.79 & 97.26 & 98.03 & 97.93 \\
k-NN \cite{ibrahim2021multilayer} & 93.00 & 90.00 & 92.20 & 91.51 \\
SVM-RBF \cite{ibrahim2021multilayer} & 85.03 & 84.87 & 86.95 & 84.95 \\
MLP \cite{ibrahim2021multilayer} & 81.95 & 86.91 & 84.58 & 83.88 \\
bot-DL \cite{10.1002/nem.2039} & 96.60 & 96.30 & 96.80 & 96.40 \\
SVM \cite{taheri2018leveraging} & 80.73 & 98.07 & 83.15 & 88.56 \\
NB \cite{10593100} & 78.00 & 79.00 & 76.00 & 77.00 \\
LR \cite{10593100} & 85.00 & 89.00 & 86.00 & 87.00 \\
AdaBoost \cite{10593100} & 90.00 & 95.00 & 94.00 & 94.00 \\
XGBoost \cite{10593100} & 96.00 & 98.00 & 98.00 & 98.00 \\
\rowcolor{green!20} \textbf{LENS-XAI Teacher} & \textbf{98.42} & \textbf{98.43} & \textbf{98.42} & \textbf{98.42} \\
\rowcolor{green!20} \textbf{LENS-XAI Student} & \textbf{98.12} & \textbf{98.14} & \textbf{98.12} & \textbf{98.12} \\
\hline
\bottomrule
\end{tabular}
}
\end{table}
 
\begin{figure}[ht]
    \centering
    \includegraphics[width=0.9\textwidth]{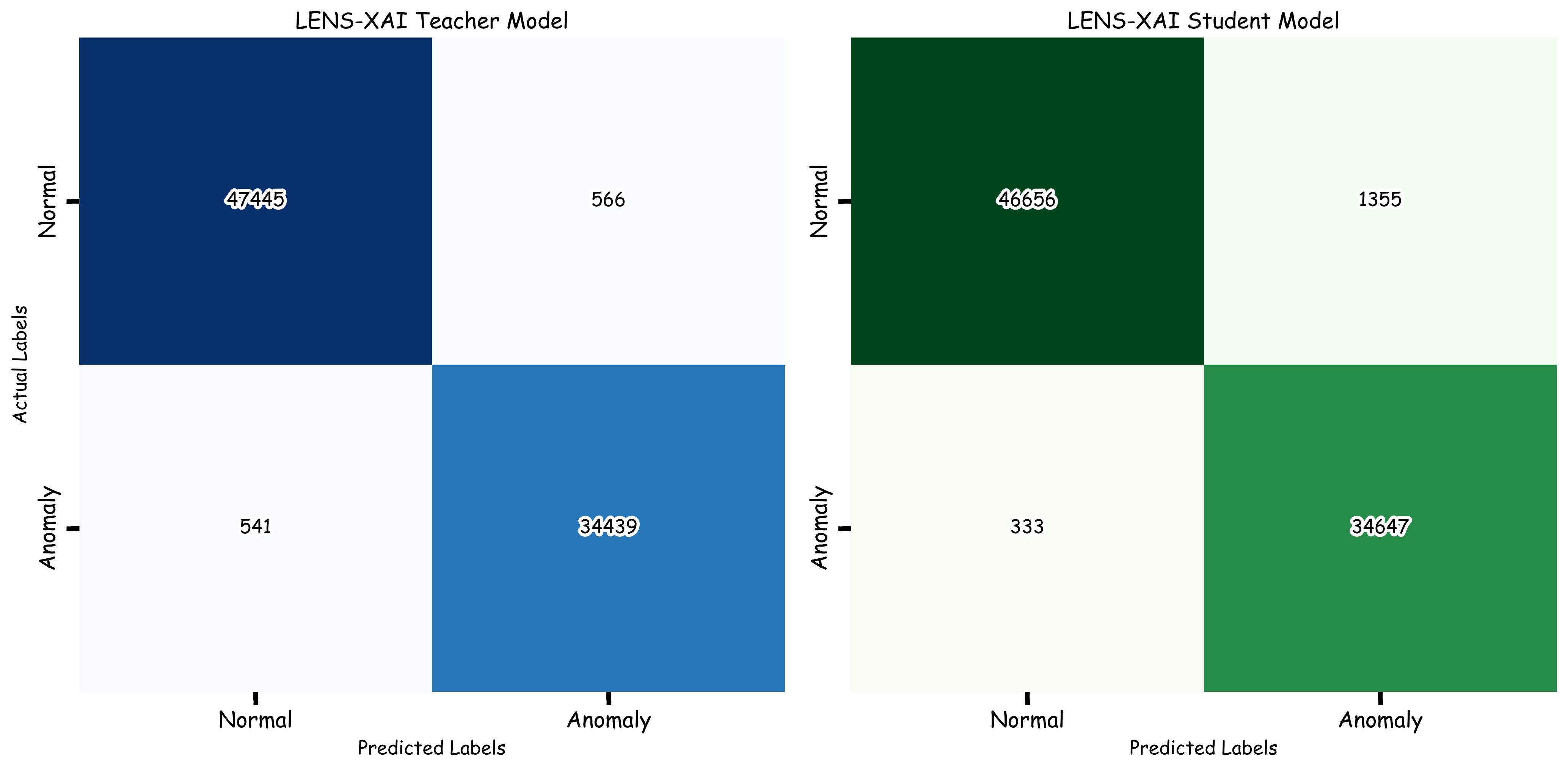}
    \vspace{-0,5em}
    \caption{Confusion matrices of the LENS-XAI framework on the CTU-13 dataset: \textit{(Left)} Teacher  and \textit{(Right)} Student models, showcasing effective differentiation between normal and anomaly classes.}
    \label{fig:CTU-13_confusion}
\end{figure}

Figure~\ref{fig:CTU-13_confusion} presents the confusion matrices of the LENS-XAI framework for both the \textit{Teacher} and \textit{Student} models on the CTU-13 dataset, illustrating their capabilities in distinguishing between normal and anomaly classes. The \textit{Teacher} model demonstrates robust performance, misclassifying only 566 normal instances as anomalies and 541 anomalies as normal, ensuring high classification accuracy. Similarly, the \textit{Student} model achieves strong results, with only 1,355 normal instances and 333 anomaly instances misclassified. These results underline the effectiveness of the LENS-XAI framework in accurately classifying network traffic while maintaining minimal misclassification rates.

\vspace{-1em}

\subsubsection{Analysis on NSL-KDD Dataset}
%Tables~\ref{tab:nsl_kdd_modelss} and~\ref{tab:nsl_kdd_binary} 
Tables~\ref{tab:nsl_kdd_modelss} 
comprehensively compares the proposed \textit{LENS-XAI} framework with state-of-the-art ML models on the NSL-KDD dataset, evaluating multi-class tasks. 

\begin{table}[h!]
\centering
\caption{Comparative performance metrics for multi-class classification on the NSL-KDD dataset.}
\vspace{0.5em}
\resizebox{\textwidth}{!}{%
\label{tab:nsl_kdd_modelss}
\begin{tabular}{@{}lcccc@{}}
\toprule
\hline
\textbf{Model} & \textbf{Accuracy (\%)} & \textbf{Precision (\%)} & \textbf{Recall (\%)} & \textbf{F1-Score (\%)} \\ 
\midrule
\hline
STL-IDS \cite{al2018deep} & 80.48 & 93.92 & 68.28 & 79.08 \\
ANN \cite{ingre2015performance} & 79.90 & N/A & N/A & N/A \\
DNN \cite{javaid2016deep} & 79.10 & 83.00 & 68.60 & 75.76 \\
AlertNet \cite{vinayakumar2019deep} & 78.50 & 81.00 & 78.50 & 76.50 \\
RNN-IDS \cite{yin2017deep} & 81.29 & N/A & 97.09 & N/A \\
RNN \cite{wu2018novel} & 81.29 & N/A & 69.73 & N/A \\
CNN \cite{wu2018novel} & 79.48 & N/A & 68.66 & N/A \\
MDNN \cite{altwaijry2019deep} & 77.55 & 81.32 & 77.55 & 75.43 \\
MCNN \cite{al2021convolutional} & 81.10 & 83.00 & 80.00 & 80.00 \\
MCNN-DFS \cite{al2021convolutional} & 81.44 & 81.00 & 84.00 & 80.00 \\
Naive Bayes \cite{al2021convolutional} & 72.73 & 76.10 & 71.90 & 77.00 \\
J48 \cite{al2021convolutional} & 74.99 & 79.60 & 75.00 & 71.10 \\
Random Forest \cite{al2021convolutional} & 76.45 & 82.10 & 75.90 & 72.50 \\
Bagging \cite{al2021convolutional} & 74.83 & 78.30 & 74.80 & 71.60 \\
Adaboost \cite{al2021convolutional} & 66.43 & N/A & 66.00 & N/A \\
\rowcolor{green!20} \textbf{LENS-XAI Teacher} & \textbf{98.66} & \textbf{98.64} & \textbf{98.67} & \textbf{98.63} \\
\rowcolor{green!20} \textbf{LENS-XAI Student} & \textbf{99.34} & \textbf{98.47} & \textbf{98.49} & \textbf{98.44} \\
\hline
\bottomrule
\end{tabular}
}
\end{table}

\begin{comment}

\begin{table}[h!]
\centering
\caption{Evaluation metrics for binary classification on the NSL-KDD dataset.}
\vspace{0.5em}
\resizebox{0.95\textwidth}{!}{%
\label{tab:nsl_kdd_binary}
\begin{tabular}{@{}lcccc@{}}
\toprule
\hline
\textbf{Model} & \textbf{Accuracy (\%)} & \textbf{Precision (\%)} & \textbf{Recall (\%)} & \textbf{F1-Score (\%)} \\ 
\midrule
\hline
\rowcolor{green!20} \textbf{LENS-XAI Teacher} & \textbf{91.62} & \textbf{91.64} & \textbf{91.62} & \textbf{91.61} \\
\rowcolor{green!20} \textbf{LENS-XAI Student} & \textbf{90.89} & \textbf{91.01} & \textbf{90.89} & \textbf{90.87} \\
\hline
\bottomrule
\end{tabular}
}
\end{table}

\end{comment}

\paragraph{\textbf{Multi-class Classification Results}}
\begin{itemize}
    \item \textbf{Accuracy}: The \textit{LENSXAI Student} achieved the highest accuracy of \textbf{99.34\%}, surpassing all competing models, including RNN-IDS (81.29\%) \cite{yin2017deep} and MCNN-DFS (81.44\%) \cite{al2021convolutional}. The \textit{Teacher} model also demonstrated remarkable accuracy at \textbf{98.66\%}, showcasing the efficacy of the knowledge transfer process.
    
    \item \textbf{Precision and Recall}: With a precision of \textbf{98.47\%} and recall of \textbf{98.49\%}, the \textit{Student} model outperformed traditional approaches like Naive Bayes (76.1\% precision) and Random Forest (75.9\% recall) \cite{al2021convolutional}. The \textit{Teacher} model closely followed, achieving precision and recall values of \textbf{98.64\%} and \textbf{98.67\%}, respectively.

    \item \textbf{F1-Measure}: Both \textit{Teacher} (\textbf{98.63\%}) and \textit{Student} (\textbf{98.44\%}) models maintained balanced performance, outperforming deep learning methods like MCNN-DFS (\textbf{80.00\%}) \cite{al2021convolutional} and STL-IDS (\textbf{79.08\%}) \cite{al2018deep}.
\end{itemize}

\begin{comment}

\paragraph{\textbf{Binary Classification Results}}
\begin{itemize}
    \item \textbf{Accuracy}: For binary classification, the \textit{LENSXAI Teacher} achieved an accuracy of \textbf{91.62\%}, while the \textit{Student} model followed with \textbf{90.89\%}. These results demonstrate the framework’s adaptability for simplified tasks without compromising performance.

    \item \textbf{Precision and Recall}: The \textit{Teacher} model attained precision and recall scores of \textbf{91.64\%} and \textbf{91.62\%}, respectively, slightly ahead of the \textit{Student} model with \textbf{91.01\%} precision and \textbf{90.89\%} recall.

    \item \textbf{F1-Measure}: Both models showcased balanced performance with F1-measures of \textbf{91.61\%} (\textit{Teacher}) and \textbf{90.87\%} (\textit{Student}), ensuring reliability in real-world binary classification scenarios.
\end{itemize}

\end{comment}

\paragraph{\textbf{Performance Analysis Across Attack Types}}
Table~\ref{tab:eval_hcrll} and the confusion matrices demonstrate the strong performance of the LENS-XAI framework, particularly the \textit{Teacher} model, which consistently outperforms the \textit{Student} model. For frequent attacks like \textbf{DoS} and \textbf{Probe}, the \textit{Teacher} achieved near-perfect accuracy (\textbf{99.50\%} and \textbf{98.61\%}), while the \textit{Student} closely followed with slight increases in misclassifications. Both models performed well on \textbf{Normal} traffic, maintaining accuracies above \textbf{98\%}. However, rare attacks like \textbf{U2R} remain challenging due to class imbalance, with the \textit{Teacher} achieving \textbf{37.14\%} accuracy compared to \textbf{23.80\%} for the \textit{Student}. In general, the LENS-XAI framework ensures high classification performance and scalability, with minimal misclassifications for major attack types.

\begin{table*}[t]
\centering 
\caption{Performance metrics comparison for the LENS-XAI framework on the NSL-KDD dataset across various attack types.}
\label{tab:eval_hcrll}
\vspace{0.5em}
\begin{small}
\resizebox{\textwidth}{!}{%
\begin{tabular}{lcccc}
\toprule
\hline

\textbf{Attack Type} & \textbf{Accuracy (\%)} & \textbf{Precision (\%)} & \textbf{Recall (\%)} & \textbf{F1 Score (\%)} \\ 
\midrule

\rowcolor{green!20}
DoS (Teacher) & \textbf{99.50} & \textbf{99.68} & \textbf{99.51} & \textbf{99.60} \\ 
\rowcolor{green!20}
DoS (Student) & \textbf{99.59} & \textbf{99.46} & \textbf{99.60} & \textbf{99.53} \\ 
\hline

\rowcolor{green!20}
Normal (Teacher) & \textbf{98.31} & \textbf{99.01} & \textbf{98.32} & \textbf{98.66} \\ 
\rowcolor{green!20}
Normal (Student) & \textbf{98.05} & \textbf{98.78} & \textbf{98.05} & \textbf{98.41} \\ 
\hline

\rowcolor{green!20}
Probe (Teacher) & \textbf{98.61} & \textbf{100.0} & \textbf{100.0} & \textbf{100.0} \\ 
\rowcolor{green!20}
Probe (Student) & \textbf{98.21} & \textbf{96.96} & \textbf{98.22} & \textbf{97.59} \\ 
\hline

U2R (Teacher) & 37.14 & 63.93 & 37.14 & 46.99 \\ 
U2R (Student) & 23.80 & 55.56 & 23.81 & \textbf{33.33} \\ 
\hline

R2L (Teacher) & 89.15 & 78.98 & 89.15 & 83.76 \\ 
R2L (Student) & 85.46 & 77.61 & 85.47 & 81.35 \\ 
\hline

\end{tabular}%
}
\end{small}
\end{table*}

\begin{figure}[ht]
    \centering
    \includegraphics[width=\textwidth]{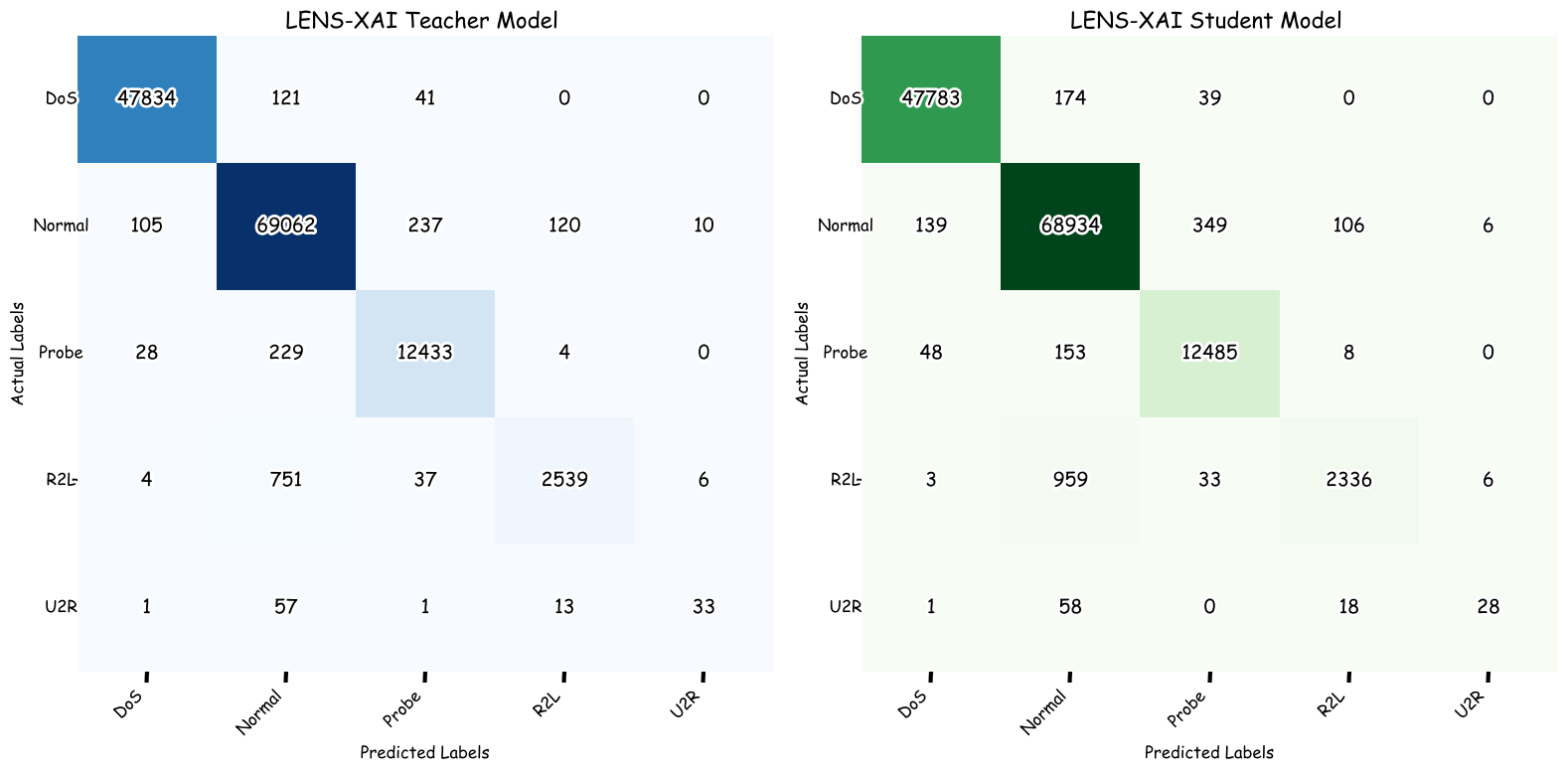}
    \caption{Confusion matrices of the LENS-XAI framework on the NSL-KDD dataset: \textit{(Left)} Teacher  and \textit{(Right)} Student models, showcasing effective differentiation between five classes.}
    \label{fig:NSL-KDDBmulti_confusion}
\end{figure}

\paragraph{\textbf{Confusion Matrix Analysis}} Figure~\ref{fig:NSL-KDDBmulti_confusion} illustrates the confusion matrices for the LENS-XAI framework, highlighting the performance of the \textit{Teacher} and \textit{Student} models on the NSL-KDD dataset for multiclass classification. The \textit{Teacher} model demonstrates exceptional accuracy with minimal misclassifications, such as 121 DoS and 237 Normal instances. The \textit{Student} model closely mirrors the \textit{Teacher}'s performance, correctly classifying 47,783 DoS and 68,934 Normal instances, showcasing its efficiency and robustness in knowledge distillation. 

%In addition, for binary classification, the \textit{Teacher} model accurately classifies 14,349 normal and 12,627 anomaly instances, while the \textit{Student} model closely follows, identifying 14,306 normal and 12,663 anomalies, with minimal misclassifications in both cases.

\begin{comment}

\begin{figure}[ht]
    \centering
    \includegraphics[width=0.9\textwidth]{nslkddbinary.png}
    \caption{Confusion matrices of the LENS-XAI framework on the NSL-KDD dataset: (Left) Teacher  and (Right) Student models, showcasing effective differentiation between normal and anomaly classes.}
    \label{fig:NSL-KDDBinary_confusion}
\end{figure}

\end{comment}

\subsection{Model Complexity and Inference Time}

\subsubsection{Multi-Class Classification Analysis}

Table \ref{tab:dataset_comparison_multiclass} provides a detailed comparison of the \textit{Student} and \textit{Teacher} models in terms of the number of parameters and inference time per batch for multi-class classification across three benchmark datasets: UKM20, Edge-IIoTset, and NSL-KDD. 

\begin{itemize}
    \item \textbf{Model Complexity:} 
        The Teacher model exhibits a significantly higher parameter count compared to the Student model across all datasets. For instance, the Teacher model has 13,065 parameters for UKM20, while the Student model only requires 4,489 parameters, representing a notable reduction in computational complexity. Similarly, for Edge-IIoTset, the parameter count is 22,415 for the Teacher model, compared to 9,167 for the Student model.
        
    \item \textbf{Inference Time:} 
        Despite its reduced parameter count, the Student model demonstrates competitive inference times. For example, the inference time per batch for UKM20 is 11.92 ms for the Student model versus 13.41 ms for the Teacher model. The Edge-IIoTset dataset shows a similar trend, with the Student model achieving an inference time of 29.77 ms, compared to 34.22 ms for the Teacher model.
\end{itemize}

\vspace{-1em}
\begin{table}[h!]
\centering
\caption{Comparison of model complexity and inference time per batch for multi-class classification scenarios.}
\vspace{0.5em}
\resizebox{\textwidth}{!}{%
\label{tab:dataset_comparison_multiclass}
\begin{tabular}{@{}lcc|cc@{}}
\toprule
\hline
\textbf{Dataset} & \multicolumn{2}{c|}{\textbf{Student Model}} & \multicolumn{2}{c}{\textbf{Teacher Model}} \\ \hline 
\hline
 & \textbf{Parameters} & \textbf{Inference Time (ms)} & \textbf{Parameters} & \textbf{Inference Time (ms)} \\ 
\midrule
\hline
 UKM20 & 4,489 & 11.92 & 13,065 & 13.41 \\
 \hline
Edge-IIoTset & 9,167 & 29.77 & 22,415 & 34.22 \\ 
\hline
 NSL-KDD & 8,197 & 28.00 & 24,325 & 31.00 \\ 
 \hline
\bottomrule
\end{tabular}
}
\end{table}

\subsubsection{Binary Classification Analysis}
Table \ref{tab:dataset_comparison_binary} outlines the comparison between the  \textit{Student} and  \textit{Teacher} models for binary classification tasks across four datasets: CTU-13, UKM20, Edge-IIoTset, and NSL-KDD.

\begin{itemize}
    \item \textbf{Model Complexity:} 
        The parameter count remains consistently lower for the Student model. For instance, the Student model for CTU-13 has 4,258 parameters, whereas the Teacher model utilizes 12,610 parameters. For Edge-IIoTset, the parameter count is 4,225 for the Student model and 12,545 for the Teacher model.
        
    \item \textbf{Inference Time:} 
        The Student model achieves superior inference times in binary classification tasks. For example, the inference time per batch for CTU-13 is 5 ms for the Student model, compared to 7 ms for the Teacher model. For Edge-IIoTset, the Student model achieves an inference time of 26.77 ms, whereas the Teacher model records 30.25 ms.
\end{itemize}

\vspace{-1em}
\begin{table}[h!]
\centering
\caption{Comparison of model complexity and inference time per batch for binary classification scenarios.}
\vspace{0.5em}
\resizebox{\textwidth}{!}{%
\label{tab:dataset_comparison_binary}
\begin{tabular}{@{}lcc|cc@{}}
\toprule
\hline
\textbf{Dataset} & \multicolumn{2}{c|}{\textbf{Student Model}} & \multicolumn{2}{c}{\textbf{Teacher Model}} \\ 

\hline
\hline
 & \textbf{Parameters} & \textbf{Inference Time (ms)} & \textbf{Parameters} & \textbf{Inference Time (ms)} \\ 
\midrule
CTU-13 & 4,258 & 5.00 & 12,610 & 7.00 \\ 
\hline
UKM20 & 4,258 & 5.00 & 12,610 & 9.00 \\ 
\hline
Edge-IIoTset & 4,225 & 26.77 & 12,545 & 30.25 \\ 
\hline
NSL-KDD & 4,225 & 8.44 & 12,545 & 9.31 \\ 
\hline
\bottomrule
\end{tabular}
}
\end{table}

These results highlight the computational efficiency of the Student model, making it a suitable choice for resource-constrained environments without a substantial sacrifice in performance.

\subsection{Variable Attribution-Based Explainability Analysis}
\label{sec:explainability_evaluation}

To uncover the inner workings of the LENS-XAI model and ensure transparency, we implemented a variable attribution-based explainability approach, detailed in Algorithm~\ref{alg:variable_attribution}. This method dissects model predictions by quantifying the contributions of individual features and latent variables, guaranteeing local accuracy. By applying this approach to datasets such as CTU-13, UKM20, NSL-KDD, and Edge-IIoTset, we gained critical insights into the factors driving model decisions. The results, presented in the subsequent sub-sections, showcase its effectiveness in binary classification for the CTU-13 dataset and both binary and multi-class tasks for the UKM20 dataset. This analysis not only highlights key predictive features but also validates the method's utility in enhancing transparency and trust for intrusion detection systems across diverse classification scenarios.

\subsubsection{Binary Classification Results on CTU-13 Dataset}
\begin{figure}[ht]
    \centering
    \includegraphics[width=1\textwidth]{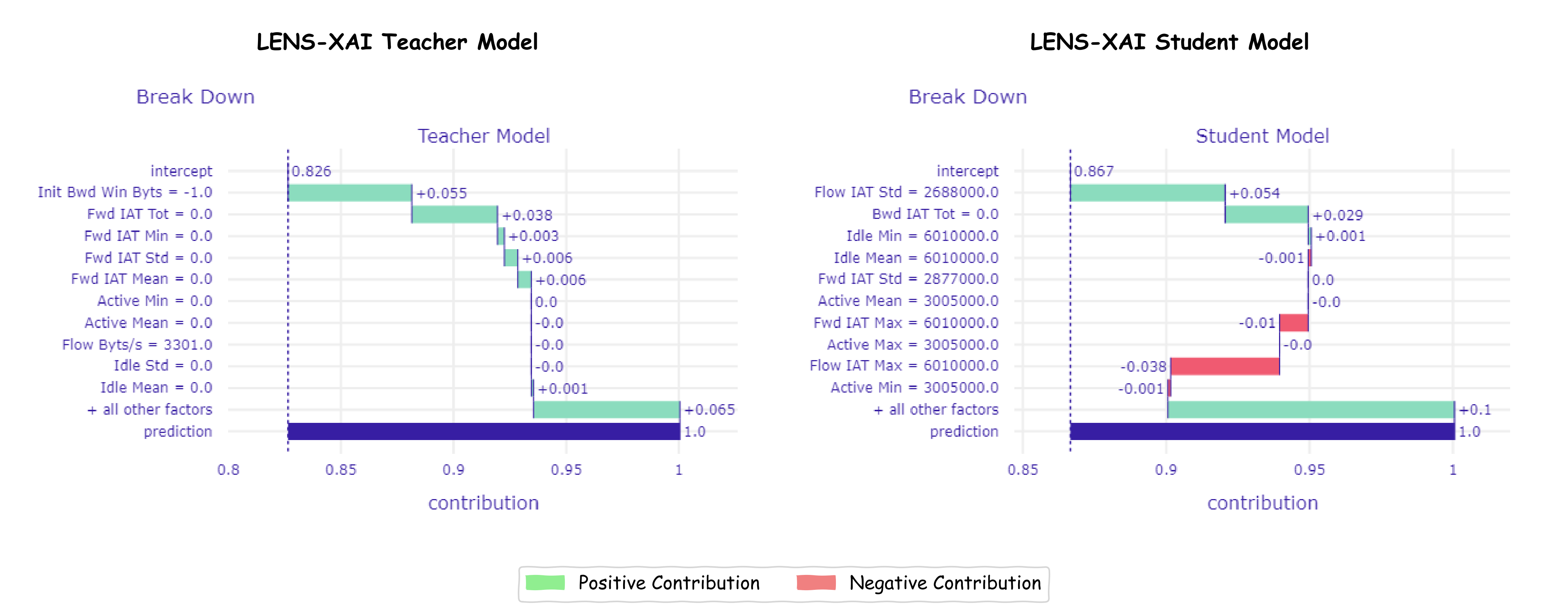}
    \vspace{-1.2em}
    \caption{Variable attribution-based explainability results for the CTU-13 dataset, highlighting key feature contributions for \textit{(Left)} Teacher and \textit{(Right)} Student models.}
\label{fig:CTU_XAI_Results}
\end{figure}

Figure~\ref{fig:CTU_XAI_Results} compares the explainability outcomes of the \emph{Teacher} and \emph{Student} models for detecting anomalies in the CTU-13 dataset under binary classification. Both models demonstrate strong predictive capability, yet they prioritize different latent dimensions and corresponding original features in their decision-making processes.

\paragraph{\textbf{Explainability Insights from the Teacher Model}}
A detailed breakdown analysis of the Teacher model (Table~\ref{tab:teacher_feature_breakdown}) indicates a final predicted value of approximately \(0.999984\) for the sample under examination. The initial intercept (\(0.419964\)) serves as the baseline probability before feature contributions are considered. Several latent dimensions—mapped to features such as \textsc{Flow Duration} and \textsc{Fwd IAT Tot}—provide positive contributions, gradually elevating the prediction closer to \(1.0\). Notably, latent dimension \(15\) (\textsc{Flow IAT Mean}) and \(1\) (\textsc{Tot Fwd Pkts}) offer significant boosts to the prediction, underlining their critical role in detecting malicious traffic. Conversely, certain dimensions (e.g., mapped to \textsc{Bwd IAT Max}) exhibit negative or minimal contributions, reflecting how the model discounts these factors for this particular instance.

By linking each latent dimension to the most influential original feature, security analysts gain insight into \emph{why} the Teacher model flags a flow as anomalous. For instance, high values in \textsc{Fwd IAT Tot} or \textsc{Flow IAT Mean} suggest irregular timing patterns, consistent with potential malicious behaviors. This transparent mapping helps validate the model’s decisions and facilitates targeted countermeasures.

\begin{table}[h!]
\centering
\caption{Breakdown of feature contributions in the Teacher model for CTU-13 dataset.}
\vspace{0.5em}
\label{tab:teacher_feature_breakdown}
\resizebox{0.85\textwidth}{!}{%
\begin{tabular}{@{}lcc@{}}
\toprule
\hline
\textbf{Mapped Feature}         & \textbf{Contribution} & \textbf{Cumulative Prediction} \\ 
\midrule
\hline
Intercept                       & +0.419964              & 0.419964                        \\
\textsc{Flow IAT Mean}                   & +0.255964              & 0.675928                        \\
\textsc{Tot Fwd Pkts}                    & +0.122159              & 0.798087                        \\
\textsc{Flow Duration}                   & +0.018376              & 0.816463                        \\
\textsc{Fwd Pkt Len Max}                 & +0.034082              & 0.850545                        \\
\textsc{Flow IAT Std}                    & +0.002265              & 0.852810                        \\
\textsc{TotLen Fwd Pkts}                 & +0.022064              & 0.874874                        \\
\textsc{TotLen Bwd Pkts}                 & -0.003403             & 0.871471                        \\
\textsc{Bwd IAT Tot}                     & -0.028823             & 0.842647                        \\
\textsc{Fwd IAT Tot}                     & +0.011063              & 0.853711                        \\
\textsc{Flow Byts/s}                     & -0.067472             & 0.786239                        \\
\textsc{Flow Pkts/s}                     & +0.050935              & 0.837174                        \\
\textsc{Bwd Header Len}                  & +0.028890              & 0.866064                        \\
\textsc{Fwd Header Len}                  & +0.044160              & 0.910225                        \\
\textsc{Bwd Pkt Len Std}                 & +0.009772              & 0.919996                        \\
\textsc{Fwd Pkt Len Std}                 & -0.007574             & 0.912422                        \\
\textsc{Fwd IAT Mean}                    & +0.014429              & 0.926851                        \\
\textsc{Tot Bwd Pkts}                    & -0.063384             & 0.863467                        \\
\textsc{Flow IAT Max}                    & -0.009915             & 0.853552                        \\
\textsc{Bwd Pkt Len Mean}                & +0.050842              & 0.904394                        \\
\textsc{Fwd Header Len}                  & -0.005207             & 0.899187                        \\
\textsc{Flow Duration}                   & -0.004680             & 0.894508                        \\
\textsc{Flow Byts/s}                     & +0.016531              & 0.911038                        \\
\textsc{Flow Pkts/s}                     & -0.004826             & 0.906212                        \\
\textsc{Bwd IAT Mean}                    & -0.080131             & 0.826081                        \\
\textsc{Fwd IAT Std}                     & +0.115819              & 0.941900                        \\
\textsc{Flow Byts/s}                     & +0.004855              & 0.946755                        \\
\textsc{Fwd Pkt Len Std}                 & +0.008506              & 0.955260                        \\
\textsc{Flow Duration}                   & +0.008806              & 0.964066                        \\
\textsc{Bwd Header Len}                  & -0.000204             & 0.963862                        \\
\textsc{Flow IAT Std}                    & +0.012679              & 0.976540                        \\
\textsc{Bwd Pkt Len Mean}                & +0.011868              & 0.988408                        \\
\textsc{Bwd IAT Tot}                     & +0.011576              & 0.999984                        \\
\hline
\bottomrule
\end{tabular}%
}
\end{table}

\paragraph{\textbf{Explainability Insights from the Student Model}}
The Student model, while aiming to reproduce the Teacher’s performance at lower computational cost, reveals a slightly different attribution pattern. Its mean predicted probability (\(0.425\)) is comparable to the Teacher’s (\(0.42\)), indicating that it has learned a similarly discriminative latent space. Through the same breakdown procedure, the Student model also emphasizes a core set of latent dimensions mapped to features like \textsc{Flow Duration}, \textsc{Bwd IAT Std}, and \textsc{Fwd IAT Mean}; however, it often distributes its attributions more broadly. For instance, \textsc{Flow IAT Std} can exhibit substantial positive contributions, hinting that variability in packet timing is a decisive indicator for intrusion detection. In certain cases, negative attributions (\textit{e.g.}, mapped to \textsc{Fwd IAT Max}) signal how the Student model can down-weight specific timing-related variables that the Teacher model might consider more crucial. Despite these nuanced differences, the Student model maintains high accuracy on the CTU-13 dataset. This consistency of essential features—coupled with some redistribution of attributions—demonstrates effective knowledge transfer. Moreover, by retaining interpretability, the Student model remains transparent regarding its learned decision boundaries, further ensuring trustworthiness in operational environments.

\subsubsection{Binary Classification Results on UKM20 Dataset}
\begin{figure}[ht]
    \centering
    \includegraphics[width=1\textwidth]{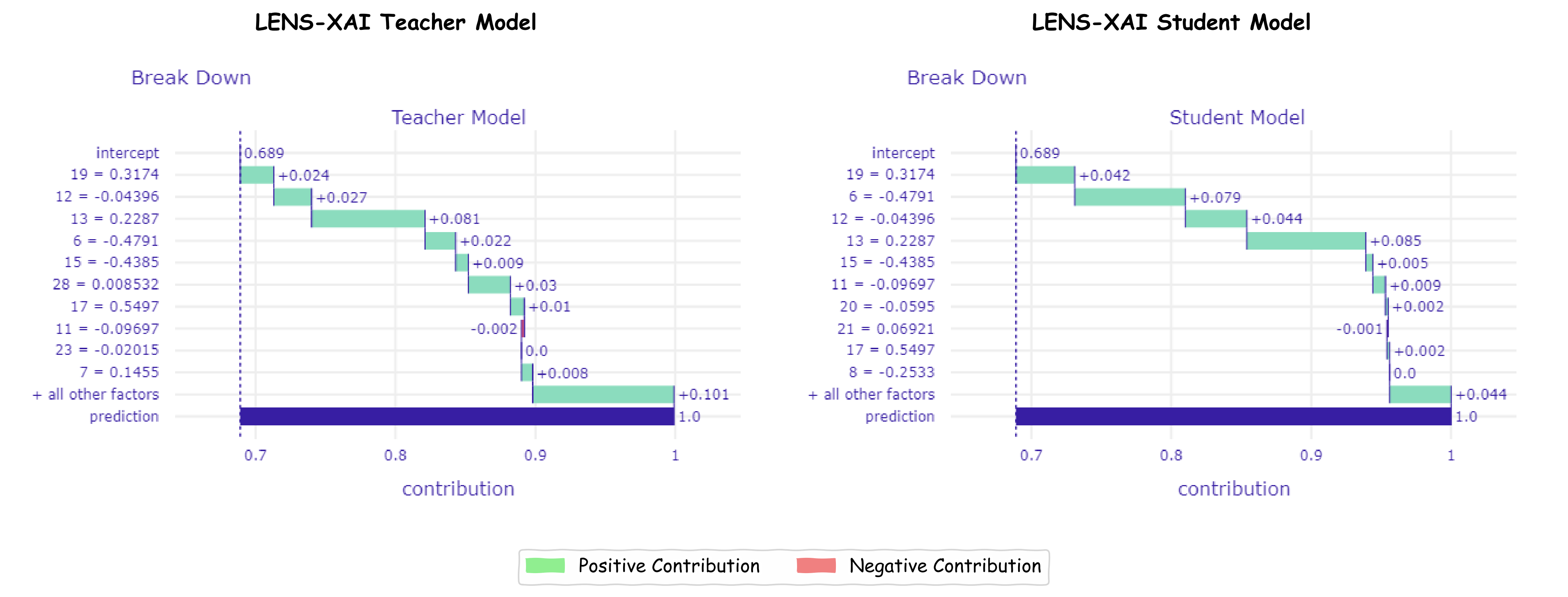}
    \vspace{-1.2em}
    \caption{Variable attribution-based explainability results for the UKM20 dataset (binary classification), highlighting key feature contributions for \textit{(Left)} Teacher and \textit{(Right)} Student models.}
\label{fig:UKM20_XAI_Results}
\end{figure}

Figure~\ref{fig:UKM20_XAI_Results} illustrates the explainability outcomes of the \emph{Teacher} and \emph{Student} models for binary classification of anomalies in the UKM20 dataset. Both models exhibit robust predictive performance, with distinct patterns of feature emphasis and latent dimension prioritization, highlighting their unique strategies for identifying and leveraging critical factors in decision-making.

\paragraph{\textbf{Explainability Insights from the Teacher Model}}
The Teacher model achieved a final predicted value of approximately \(0.9990\), with an initial baseline probability (intercept) of \(0.689\). Key features such as \textsc{Source Bytes}, \textsc{Flow Duration}, and \textsc{Packet Interarrival Times} made significant positive contributions, driving the prediction closer to \(1.0\).
For example:
\begin{itemize}
    \item \textsc{Source Bytes} contributed \(+0.0418\), indicating high data transfer activity—a characteristic often linked to potential malicious behavior.
    \item \textsc{Flow Duration} added \(+0.0794\), suggesting irregular timing patterns, which are crucial for identifying anomalies.
\end{itemize}

Conversely, features like \textsc{Backward Packet Timing Variability} had minimal or negative contributions (\(-0.0020\)), demonstrating the Teacher model's ability to down-weight less influential factors. This selective feature attribution enhances the model's precision in flagging anomalies. By mapping each latent dimension to influential features, security analysts can better understand the model's rationale.

\paragraph{\textbf{Explainability Insights from the Student Model}}
The Student model, designed to replicate the Teacher’s performance with lower computational demands, produced a final predicted value of \(0.999971\). While the Student model prioritized many of the same key features as the Teacher model, it exhibited a broader attribution pattern. 
For example:
\begin{itemize}
    \item \textsc{Source Bytes} (\(+0.0418\)) and \textsc{Flow Duration} (\(+0.0794\)) remained prominent contributors.
    \item Additional features like \textsc{Packet Timing Variability} and \textsc{Destination Bytes} received notable attributions, reflecting a more distributed emphasis on various characteristics of the dataset.
\end{itemize}

Interestingly, the Student model de-emphasized certain features such as \textsc{Average Packet Sizes} (\(-0.0149\)), which were down-weighted when deemed less relevant. This redistribution of attributions demonstrates the effectiveness of knowledge distillation, allowing the Student model to maintain high performance while reducing computational complexity.

\subsubsection{Multi-Class Classification Results on UKM20 Dataset}
\vspace{0.5em}
\paragraph{\textbf{Explainability Insights from the Teacher Model}}
The Teacher model achieved a final predicted value of approximately \(0.999\) with an initial baseline probability (intercept) of \(0.034\). Noteworthy contributions to the prediction came from features such as:
\begin{itemize}
    \item \textsc{Source Bytes}, which contributed \(+0.0512\), signifying data transfer activity relevant to multi-class distinctions.
    \item \textsc{Flow Duration}, adding \(+0.0507\), which indicates temporal patterns significant for differentiating classes.
    \item \textsc{Backward Packet Timing Variability}, which contributed minimally or negatively (\(-0.0002\)), showing its lower importance for this task.
\end{itemize}

These contributions highlight the Teacher model's ability to identify and focus on critical features, ensuring precision in multi-class predictions. By mapping latent dimensions to influential features, the model offers transparency and interpretability for multi-class classification, enabling analysts to validate its predictions effectively.

\paragraph{\textbf{Explainability Insights from the Student Model}}
The Student model, designed to replicate the Teacher’s performance with reduced computational complexity, produced a final predicted value of \(0.956\). While prioritizing several key features similar to the Teacher model, the Student model distributed its attributions more broadly. Key insights include:
\begin{itemize}
    \item \textsc{Source Bytes} contributed \(+0.0323\), reinforcing its significance across both models.
    \item \textsc{Flow Duration} added \(+0.0098\), demonstrating its continued importance for classification.
    \item Features such as \textsc{Packet Interarrival Times} and \textsc{Destination Bytes} showed broader attribution, reflecting the Student model's distributed emphasis.
    \item Negative contributions, like \textsc{Average Packet Sizes} (\(-0.0015\)), illustrate the Student model's nuanced feature weighting.
\end{itemize}

This broader attribution pattern highlights the effectiveness of the Student model in balancing accuracy with computational efficiency. The redistribution of feature contributions demonstrates the success of knowledge distillation from the Teacher model.

\paragraph{\textbf{Comparative Analysis}}
Table~\ref{tab:ukm20_multiclass} provides a comprehensive breakdown of feature contributions for the \textit{Teacher} and \textit{Student} models on the UKM20 dataset. Key observations from the table include the following:

\begin{itemize}
    \item The \textsc{Source Bytes} and \textsc{Flow Duration} features had the most significant positive contributions for both models, underlining their critical role in detecting anomalies.
    \item The Teacher model demonstrated precision by concentrating on a limited set of highly impactful features, such as \textsc{Push Packets} (\(+0.0051\)) and \textsc{End Time} (\(+0.0050\)).
    \item The Student model exhibited a more distributed attribution pattern, with moderate contributions spread across features like \textsc{Backward Packets} (\(+0.0032\)) and \textsc{Destination Bytes} (\(+0.0068\)).
    \item Certain features, such as \textsc{Transport Protocol Flags} (\(-0.0032\) for Teacher and \(-0.0021\) for Student), contributed negatively, reflecting their reduced relevance in anomaly detection.
    \item Both models consistently identified timing-related features (\textsc{Packet Timing Variability} and \textsc{Round Trip Time (First ACK)}) as moderately significant, emphasizing the importance of temporal patterns.
\end{itemize}

These findings highlight the robustness of the LENS-XAI framework, offering interpretable and accurate predictions. The Teacher model's focus on precision ensures reliability in high-stakes environments, while the Student model’s adaptability makes it suitable for resource-constrained operational settings.

\begin{table}[h!]
\centering
\caption{Breakdown of feature contributions for multi-class classification on UKM20 dataset.}
\vspace{0.5em}
\label{tab:ukm20_multiclass}
\resizebox{1.0\textwidth}{!}{%
\begin{tabular}{@{}lcc@{}}
\toprule
\hline
\textbf{Feature}                   & \textbf{Teacher Contribution} & \textbf{Student Contribution} \\ 
\midrule
\hline
Intercept                          & \(+0.03396\)                 & \(+0.03243\)                  \\
\textsc{Source Bytes}              & \(+0.0512\)                  & \(+0.0323\)                   \\
\textsc{Flow Duration}             & \(+0.0507\)                  & \(+0.0098\)                   \\
\textsc{Backward Timing Variability} & \(-0.0002\)                & \(-0.0015\)                   \\
\textsc{Packet Timing Variability}  & \(+0.0032\)                  & \(+0.0047\)                   \\
\textsc{Destination Bytes}         & \(+0.0021\)                  & \(+0.0068\)                   \\
\textsc{Average Packet Sizes}      & \(+0.0015\)                  & \(-0.0011\)                   \\
\textsc{Urgent Bits}               & \(+0.0007\)                  & \(+0.0004\)                   \\
\textsc{Push Packets}              & \(+0.0051\)                  & \(+0.0046\)                   \\
\textsc{Transport Protocol}        & \(-0.0089\)                  & \(-0.0011\)                   \\
\textsc{Start Time}                & \(+0.0033\)                  & \(+0.0022\)                   \\
\textsc{End Time}                  & \(+0.0050\)                  & \(+0.0047\)                   \\
\textsc{Host-to-Host Count}        & \(+0.0032\)                  & \(+0.0038\)                   \\
\textsc{Round Trip Time (First ACK)} & \(+0.0011\)                & \(+0.0008\)                   \\
\textsc{Average Time Sent}         & \(+0.0029\)                  & \(+0.0026\)                   \\
\textsc{Average Time Received}     & \(+0.0031\)                  & \(+0.0028\)                   \\
\textsc{Repeated Flows}            & \(-0.0005\)                  & \(-0.0007\)                   \\
\textsc{Forward Packets}           & \(+0.0063\)                  & \(+0.0055\)                   \\
\textsc{Backward Packets}          & \(+0.0046\)                  & \(+0.0032\)                   \\
\textsc{Source TTL}                & \(+0.0055\)                  & \(+0.0049\)                   \\
\textsc{Destination TTL}           & \(+0.0068\)                  & \(+0.0056\)                   \\
\textsc{Transport Protocol Flags}  & \(-0.0032\)                  & \(-0.0021\)                   \\
\hline
\bottomrule
\end{tabular}%
}
\end{table}

\section{Discussion}
\label{sec:5-discussion}
The increasing prevalence of interconnected systems and the concurrent rise in cyber threats highlight the critical need for advanced IDS that prioritize accuracy, explainability, and scalability. The proposed \textbf{LENS-XAI} framework addresses these challenges by integrating VAEs, knowledge distillation, and variable attribution-based explainability mechanisms into a cohesive and lightweight architecture. This study provides compelling evidence, through extensive evaluations (see {Table~\ref{tab:dataset_comparison}}), of the framework's potential to safeguard resource-constrained environments such as IoT and edge computing systems.

\subsection{Key Findings and Implications}

\paragraph{\textbf{Superior Detection Performance Across Datasets}} The LENS-XAI framework demonstrated superior performance across benchmark datasets, consistently achieving high detection accuracy. As detailed in Section~\ref{sec:baseline_methods} and Table~\ref{tab:edge_iiotset_comparisonn}, the framework achieved an accuracy of 95.34\% on the Edge-IIoTset dataset with the Teacher model and 95.31\% with the Student model. On the UKM20 dataset, it achieved near-perfect accuracies of 99.92\% (Teacher) and 99.80\% (Student) as highlighted in Table~\ref{tab:ukm20_multiclasss}. Similarly, for the CTU-13 dataset, the framework outperformed other models, achieving 98.42\% accuracy (Teacher) and 98.12\% (Student), as shown in Table~\ref{tab:ml_models}. For the NSL-KDD dataset, multi-class classification accuracies of 98.66\% (Teacher) and 99.34\% (Student) were observed, outperforming other state-of-the-art methods (Table~\ref{tab:nsl_kdd_modelss}). 

These results affirm the robustness of the LENS-XAI framework in detecting a wide range of cyber threats, including frequent and rare attack types. The framework's ability to maintain low false-positive rates, as evident from confusion matrices in Figures~\ref{fig:Edge-IIoTSetMulticlass_confusion} - \ref{fig:NSL-KDDBmulti_confusion}, ensures its practical utility in real-world scenarios.

\paragraph{\textbf{Enhanced Explainability and Interpretability}} The integration of variable attribution methods provides a granular understanding of the framework's decision-making process, as discussed in Section~\ref{sec:explainability_evaluation}. The Teacher model identified critical features such as \textsc{Source Bytes} and \textsc{Flow Duration}, with significant contributions to anomaly detection (Tables~\ref{tab:teacher_feature_breakdown} and \ref{tab:ukm20_multiclass}). Visualization results in Figures~\ref{fig:CTU_XAI_Results} and \ref{fig:UKM20_XAI_Results} highlight the importance of temporal patterns in detecting anomalies. 

The Student model effectively replicated the Teacher model's interpretability, albeit with broader feature attribution patterns. This balance between explainability and computational efficiency makes LENS-XAI suitable for resource-constrained environments, such as IoT and edge devices.

\paragraph{\textbf{Lightweight Design and Scalability}} The knowledge distillation approach enabled the Student model to achieve comparable detection performance to the Teacher model while significantly reducing resource consumption, as detailed in Table~\ref{tab:ukm_model_comparisonn}. This scalability ensures that LENS-XAI can be deployed in environments with limited computational resources. For instance, The Student model achieved up to a 44\% reduction in computational overhead (average reduction of 23.47\%) while maintaining detection rates above 95\% across dataset.

\paragraph{\textbf{Practical Utility in Real-World Applications}} The comprehensive evaluation of LENS-XAI highlights its practicality for real-world deployment. The framework's ability to balance high performance, scalability, and interpretability makes it a valuable tool for cybersecurity. Furthermore, explainability analyses presented in Figures~\ref{fig:CTU_XAI_Results} and \ref{fig:UKM20_XAI_Results} reinforce stakeholder trust by offering transparent decision-making processes.

\subsection{Comparison with Existing Methods}
The \textbf{LENS-XAI} framework advances beyond traditional IDS by combining high accuracy, lightweight architecture, and enhanced interpretability. As demonstrated in Tables~\ref{tab:ml_models} and \ref{tab:nsl_kdd_modelss}, it consistently outperformed models such as RNN, CNN, and Random Forest in terms of accuracy and explainability. The use of knowledge distillation further enhances its suitability for resource-constrained environments, setting it apart from black-box approaches like bot-DL and FFS-HTTP, which lack interpretability (Table~\ref{tab:ml_models}). These comparisons underline the versatility and robustness of LENS-XAI as a state-of-the-art IDS, capable of addressing the diverse challenges of modern cybersecurity.

\subsection{Limitations}
Despite the promising results, this study acknowledges certain limitations that require further investigation:
\begin{itemize}
    \item \textbf{Dependence on Labeled Data:} The reliance on labeled datasets restricts the framework's applicability in scenarios where annotated data is scarce. Future research could explore unsupervised or semi-supervised learning techniques to mitigate this limitation.
    \item \textbf{Explainability vs. Computational Overhead:} While the inclusion of explainability mechanisms strengthens transparency, it may introduce additional computational costs. Optimizing these mechanisms for real-time applications is an important avenue for future work.
    \item \textbf{Evaluation on Dynamic Threat Profiles:} Expanding the evaluation to datasets with evolving and dynamic threat landscapes (Table~\ref{tab:dataset_comparison}) is crucial to ensure the framework's robustness against emerging cyber threats.
\end{itemize}

\section{Conclusion}
\label{sec:6-conclusion}
This research introduces a novel intrusion detection framework, LENS-XAI, which integrates lightweight design principles with enhanced interpretability and scalability, utilizing a training set comprising $10\%$ of the available data to optimize model efficiency without compromising performance. The framework employs variable attribution methods to provide transparent insights into its decision-making process while maintaining superior performance across multiple benchmark datasets. Initial preprocessing ensures data consistency, followed by effective feature extraction and dimensionality reduction to optimize model performance. The Teacher-Student architecture of LENS-XAI achieves a robust balance between high detection accuracy and computational efficiency. Notably, the framework recorded accuracies of 95.34\%, 99.92\%, 98.42\%, and 99.34\% on the Edge-IIoTset, UKM20, CTU-13, and NSL-KDD datasets, respectively, outperforming state-of-the-art models.

The analysis highlights that \textbf{LENS-XAI} not only excels in detecting frequent and complex attack types but also addresses challenges such as class imbalance and overfitting, improving generalization to unseen test cases. Furthermore, its scalability and explainability make it an ideal candidate for real-world deployment in resource-constrained environments like IoT and edge computing systems. The framework reduces computational overhead and time complexity, enhancing its practical utility for safeguarding interconnected systems. Future work could focus on addressing limitations in detecting rare attack types and exploring ensemble AI systems to further improve intrusion detection in distributed and evolving threat environments. By advancing these areas, LENS-XAI aims to strengthen its role as a next-generation cybersecurity solution.

\vspace{-0.5em}

\section*{Funding}
This research did not receive specific grants from public, commercial, or not-for-profit funding agencies.

\vspace{-1em}

\section*{Declaration of Competing Interest}
The authors affirm that they have no competing financial interests or personal relationships that could have influenced the work presented in this paper.

\vspace{-1em}

\section*{Acknowledgments}
None

\vspace{-1em}

\section*{Data Availability}
Mentioned in the manuscript.

\bibliographystyle{elsarticle-num}  %% Elsevier style for numbered references
\bibliography{references}          %% Ensure references.bib exists in the same directory

\newpage

\section*{Appendix I: Attack Types and Definitions}
\label{sec:appendix_datasets}
\appendix
\section*{1. Overview of Attack Types}
\label{sec:appendix_datasets}

\begin{itemize}
    \item \textbf{Backdoor:} A covert method to bypass standard authentication mechanisms, granting unauthorized access to a system. Often introduced via malware.

    \item \textbf{Distributed Denial of Service (DDoS):} A coordinated cyberattack that disrupts the normal functioning of a network, server, or application by overwhelming it with excessive traffic. Common subtypes include:
    \begin{itemize}
        \item \textbf{HTTP Flood:} Exploits Hypertext Transfer Protocol (HTTP) GET or POST requests to target web servers or applications.
        \item \textbf{ICMP Flood:} Overloads the target with Internet Control Message Protocol (ICMP) echo requests (commonly known as ``ping traffic").
        \item \textbf{TCP Flood:} Exploits Transmission Control Protocol (TCP) connections to deplete system resources.
        \item \textbf{UDP Flood:} Sends massive amounts of User Datagram Protocol (UDP) packets to random ports, overwhelming the target.
    \end{itemize}

    \item \textbf{Fingerprinting:} The identification of devices, systems, or applications by analyzing unique characteristics, typically used for reconnaissance purposes.

    \item \textbf{Man-In-The-Middle (MITM):} An attack where an adversary intercepts and manipulates communications between two parties without their knowledge.

    \item \textbf{Password Attacks:} Methods to obtain user passwords, including brute force, dictionary attacks, and other techniques.

    \item \textbf{Port Scanning:} A reconnaissance technique used to identify open ports and active services on a target system.

    \item \textbf{Ransomware:} A form of malicious software (malware) that encrypts the victim's data, demanding payment for decryption keys.

    \item \textbf{SQL Injection:} A database manipulation attack that exploits vulnerabilities in Structured Query Language (SQL) queries to access or alter sensitive information.

    \item \textbf{Uploading Attacks:} Exploits vulnerabilities in web applications by uploading malicious files to gain unauthorized access or disrupt systems.

    \item \textbf{Vulnerability Scanning:} Automated processes designed to detect and potentially exploit known system vulnerabilities.

    \item \textbf{Cross-Site Scripting (XSS):} Injection of malicious scripts into trusted websites to execute on a user’s browser, often for credential theft.

    \item \textbf{ARP Poisoning:} Alters Address Resolution Protocol (ARP) tables to intercept or manipulate network traffic.

    \item \textbf{BeEF HTTP Exploits:} Attacks leveraging the Browser Exploitation Framework (BeEF) to exploit vulnerabilities in web browsers.

    \item \textbf{Mass HTTP Requests:} Overwhelms servers by generating a high volume of simultaneous HTTP requests.

    \item \textbf{Metasploit Exploits:} Attacks using the Metasploit framework to exploit software or system vulnerabilities.

    \item \textbf{TCP Flood:} Overloads a target by initiating numerous TCP connections simultaneously.

    \item \textbf{UDP Data Flood:} Sends a large volume of UDP packets with random or invalid data to the target, causing disruption.

    \item \textbf{Botnet:} A network of compromised devices controlled remotely by attackers, often used for large-scale attacks such as DDoS.

    \item \textbf{Denial of Service (DoS):} An attack that disrupts system availability by overloading it with requests or malicious traffic.

    \item \textbf{Probe:} Scans networks or systems to gather information for potential exploitation.

    \item \textbf{Remote-to-Local (R2L):} An attack where a remote user gains unauthorized local access to a system.

    \item \textbf{User-to-Root (U2R):} A privilege escalation attack where an attacker gains administrative or root access to a system.
\end{itemize}

\section*{2. Dataset Accessibility}

\begin{itemize}
    \item \textbf{Edge-IIoTset Dataset:} A comprehensive and realistic cyber-security dataset for Internet of Things (IoT) and Industrial IoT (IIoT) applications. Available at: 
    \footnote{\url{https://ieee-dataport.org/documents/edge-iiotset-new-comprehensive-realistic-cyber-security-dataset-iot-and-iiot-applications}}.

    \item \textbf{UKM-IDS20 Dataset:} Focused on Unmanned Aerial Vehicle (UAV) communication security, this dataset can be accessed at:
    \footnote{\url{https://www.kaggle.com/datasets/muatazsalam/ukm-ids20}}.

    \item \textbf{CTU-13 Dataset:} Includes botnet traffic mixed with normal traffic, widely used in network security studies. Available at: 
    \footnote{\url{https://www.stratosphereips.org/datasets-ctu13}}.

    \item \textbf{NSL-KDD Dataset:} A refined version of the Knowledge Discovery in Databases (KDD) Cup 1999 dataset, aimed at reducing redundancy and imbalance. Accessible at: 
    \footnote{\url{https://www.kaggle.com/datasets/hassan06/nslkdd}}.
\end{itemize}

\end{document}